\documentclass[10pt,notitlepage]{report}
\usepackage{amsmath,amssymb} 

\usepackage{fullpage}
\usepackage{graphicx}
\usepackage[dvips]{epsfig}
\usepackage{pbox}
\usepackage{authblk}

\renewcommand{\a}{\alpha}
\renewcommand{\b}{\beta}
\newcommand{\g}{\gamma}

\renewcommand{\d}{\delta}

\newcommand{\s}{\sigma}
\newcommand{\D}{\Delta}
\renewcommand{\th}{\theta}

\renewcommand{\o}{\omega}
\renewcommand{\O}{\Omega}
\newcommand{\e}{\epsilon}
\newcommand{\ve}{\varepsilon}
\renewcommand{\l}{\lambda}
\renewcommand{\vec}[1]{\boldsymbol{#1}}

\newcommand{\ua}{\uparrow}
\newcommand{\da}{\downarrow}
\newcommand{\su}{\uparrow}
\newcommand{\sd}{\downarrow}

\renewcommand{\dag}{\dagger}

\newcommand{\lb}{\label}
\newcommand{\nn}{\nonumber}

\newtheorem{entry}{}[section]
\newcommand{\bent}[1]{\vspace*{-1cm}\hspace*{-1cm}\begin{entry}\lb{e{#1}}\rm}
\newcommand{\eent}{\end{entry}}

\newcommand{\BK}[1]{\left[#1\right]}

\newcommand{\bra}[1]{\left\langle{#1}\right|}
\newcommand{\ket}[1]{\left|{#1}\right\rangle}
\newcommand{\lr}[1]{\left\langle#1\right\rangle}

\newcommand{\bdag}{b^{\dagger}}
\newcommand{\bbbb}{b^{\phantom{\dagger}}}

\newcommand{\be}{\begin{equation}}
\newcommand{\ee}{\end{equation}}

\renewcommand{\v}[2]{\left[\begin{array}{c} #1\\#2 \end{array}\right]}
\newcommand{\m}[4]{\left[\begin{array}{cc} #1&#2\\#3&#4 \end{array}\right]}

\begin{document}

\title{Spin-orbit Coupling in Optical lattices}
\author[1]{Shizhong Zhang}
\author[2]{William S. Cole}
\author[3]{Arun Paramekanti}
\author[2]{Nandini Trivedi}
\affil[1]{Department of Physics and Center of Theoretical and Computational Physics, The University of Hong Kong, Hong Kong, China}
\affil[2]{Department of Physics, The Ohio State University, Colombus, Ohio, 43210, USA}
\affil[3]{Department of Physics, University of Toronto, Toronto M5S1A7, and Canadian Institute for Advanced Research, Toronto, Ontario, M5G 1Z8, Canada}
%\author{Shizhong Zhang\\Department of Physics and Center of Theoretical and Computational Physics, \\
%The University of Hong Kong, Hong Kong, China}
%William S. Cole\\Department of Physics, The Ohio State University, Colombus, Ohio, 43210, USA
%\affiliation{Department of Physics and Center of Theoretical and Computational Physics, The University of Hong Kong, Hong Kong, China}
%\author{William S. Cole}
%\affiliation{Department of Physics, The Ohio State University, Colombus, Ohio, 43210, USA}
%\author{Arun Paramekanti}
%\affiliation{Department of Physics, University of Toronto, Toronto M5S1A7, and Canadian Institute for Advanced Research, Toronto, Ontario, M5G 1Z8, Canada}
%\author{Nandini Trivedi}
%\affiliation{Department of Physics, The Ohio State University, Colombus, Ohio, 43210, USA}

\maketitle

\begin{abstract}
In this review, we discuss the physics of spin-orbit coupled quantum gases in optical lattices. After reviewing some relevant experimental techniques, we introduce the basic theoretical model and discuss some of its generic features. In particular, we concentrate on the interplay between spin-orbit coupling and strong interactions and show how it leads to various exotic quantum phases in both the Mott insulating and superfluid regimes. Phase transitions between the Mott and superfluid states are also discussed.
\end{abstract}

\section{Introduction}

Cold atom experiments are performed with charge-neutral atoms~\cite{Pitaevskii2003,Pethick2008}. At first sight, this would have precluded the effects of orbital magnetism, as well as spin-orbit coupling to be studied in these cold atomic gases. However, in the past few years, by using atom-light coupling (Raman lasers and shaking optical lattice), it has become possible to simulate these effects in neutral atomic sample. This provides cold atom experimentalists with the exciting opportunity to produce and investigate several paradigmatic quantum states such as the quantum Hall liquids, topological insulators and superfluids, Dirac and Weyl semimetals, as well as many other exotic phenomena that have recently been predicted for electron systems in external magnetic fields or with strong spin-orbit interaction~\cite{Hasan2010,Hasan2011,Qi2011}. What is perhaps more interesting is that this capacity would open an entire new vista for the investigation of bosonic topological states that have so far only been subjected to theoretical studies~\cite{Vishwanath2013,Chen2013,Senthil2013}. Indeed, the great tunability of cold atom systems provides an avenue to the realization of conceptually important models that may not have a natural correspondence to a condensed matter system~\cite{Esslinger2010,Windpassinger2013}, as was beautifully demonstrated in the recent implementation of Haldane's honeycomb model of a Chern insulator in a cold atom experiment~\cite{Jotzu2014}. 

As a result of this fundamental interest, there has been tremendous effort in realizing synthetic magnetic flux and spin-orbit coupling in neutral atomic gases over the past few years; for recent reviews, see refs.~\cite{Dalibard2011, Goldman2013, Galitski2014, Hui2014}. Many schemes have been proposed and several of them have been implemented in experiments. At present, there are two schemes which have accumulated the most substantial experimental success: the Raman scheme~\cite{Lin2012,Cheuk2012,Wang2012,Zhang2012,Hamner2014} and shaken optical lattices~\cite{Struck2011,Struck2012,Struck2013,Jotzu2014}. The former has been utilized to produce both synthetic magnetic flux and spin-orbit coupling, while the later has so far only been used to produce synthetic magnetic flux. However, several proposals exist in the literature using shaking optical lattices to produce spin-orbit coupling~\cite{Goldman2014,Struck2014}. In this review, we shall concentrate our attention on the experimentally implemented schemes. In Table \ref{soc_exp}, we give a summary of the experiments conducted so far on spin-orbit coupled quantum gases using Raman scheme.

%--------------------------Table--------------------------
\begin{table}[h]
\def\arraystretch{2}
\setlength{\tabcolsep}{1 em}
\begin{center}
    \caption{\label{soc_exp}Experiments on spin-orbit coupled quantum gases.}
    \begin{tabular}{ | c | c | c | c |}
    \hline
      Group & Element  & Phenomena  & Comments \\[3pt] \hline		
    NIST & $^{87}$Rb & \pbox{6 cm}{structure of BEC~\cite{Lin2012}; partial wave scattering~\cite{Williams2012}; spin hall effect~\cite{Beeler2013}; Zitterbewegung~\cite{LeBlanc2013}} & \pbox{4 cm}{harmonic trap }\\ [3pt] \hline
    USTC & $^{87}$Rb & \pbox{6 cm}{dipole oscillation~\cite{Zhang2012}; finite temperature phase diagram~\cite{Ji2014}; collective excitations~\cite{Ji2014a}} & harmonic trap \\ [3pt]\hline
    Shanxi & $^6$Li & \pbox{6 cm}{ARPES, Fermi surface transition~\cite{Wang2012}; spin-orbit coupled molecule~\cite{Fu2013}} & \pbox{2.8cm}{fermion; \\ harmonic trap }\\ [3pt]\hline
    MIT	& $^6$Li & inverse ARPES, Zeeman Lattice~\cite{Cheuk2012} & fermion\\ [3pt] \hline
    Purdue	& $^{87}$Rb & Landau-Zener transitions~\cite{Olson2014} & harmonic trap\\[3pt]\hline
     WSU	& $^{87}$Rb & \pbox{6 cm}{dynamical instability of spin-orbit BEC ~\cite{Hamner2014}, collective excitations~\cite{Khamehchi2014}} & moving optical lattice\\[3pt]\hline
    \end{tabular}
\end{center}
\def\arraystretch{1.5}
\end{table}
%--------------------------Table--------------------------

For a general overview of the subject, we refer readers to ref.~\cite{Galitski2014}. For a more complete introduction to the subject and in particular, on the experimental techniques, see refs.~\cite{Dalibard2011,Goldman2014}. Results related to spin-orbit coupled quantum gases in a harmonic trap are reviewed in ref.~\cite{Hui2014}, which concentrates mostly on the weakly interacting regime. In this review, we focus rather on the interplay between the effects of spin-orbit coupling and strong interactions. In the case of bosons, perhaps the simplest route to this regime is to load bosons in an deep optical lattice with spin-orbit coupling generated either by the Raman scheme or by shaking, as we shall review below in Sec.~\ref{sec:er}. In Sec.~\ref{sec:btm} we then move on to review the basic theoretical models that describe spin-orbit coupled bosons in the optical lattice and discuss the resulting band structure in Sec.~\ref{sec:bs}. In Sec.~\ref{sec:sip} we discuss some general themes that emerge from the interplay between spin-orbit coupling and strong interactions, specifically identifying a few of the more novel aspects. Finally, in Sec.~\ref{sec:prospect},  we conclude our review and offer some perspectives on the subject.

\section{Experimental realizations}
\label{sec:er}
So far, several ways of generating spin-orbit coupling in optical lattices have been proposed and, in some cases, implemented. Building on the Raman scheme which realizes the effect of spin-orbit coupling in the uniform system, an extra pair of lattice beams can be added. This has been achieved in a recent experiment~\cite{Hamner2014} with a moving optical lattice. Other schemes for generating spin-orbit coupling include Raman-assisted tunneling~\cite{Kennedy2013} and shaking the optical lattice~\cite{Goldman2014,Struck2014}. A particularly interesting idea is the so-called ``Zeeman" lattice, in which the spin-orbit coupling and optical lattice are generated at the same time by a combination of radio-frequency beams and Raman beams~\cite{Garcia2012,Cheuk2012}. We shall discuss each of these techniques in turn.

\subsection{Raman scheme with an optical lattice}
\begin{figure}[ht]
\begin{center}
\includegraphics[width = 0.8\textwidth]{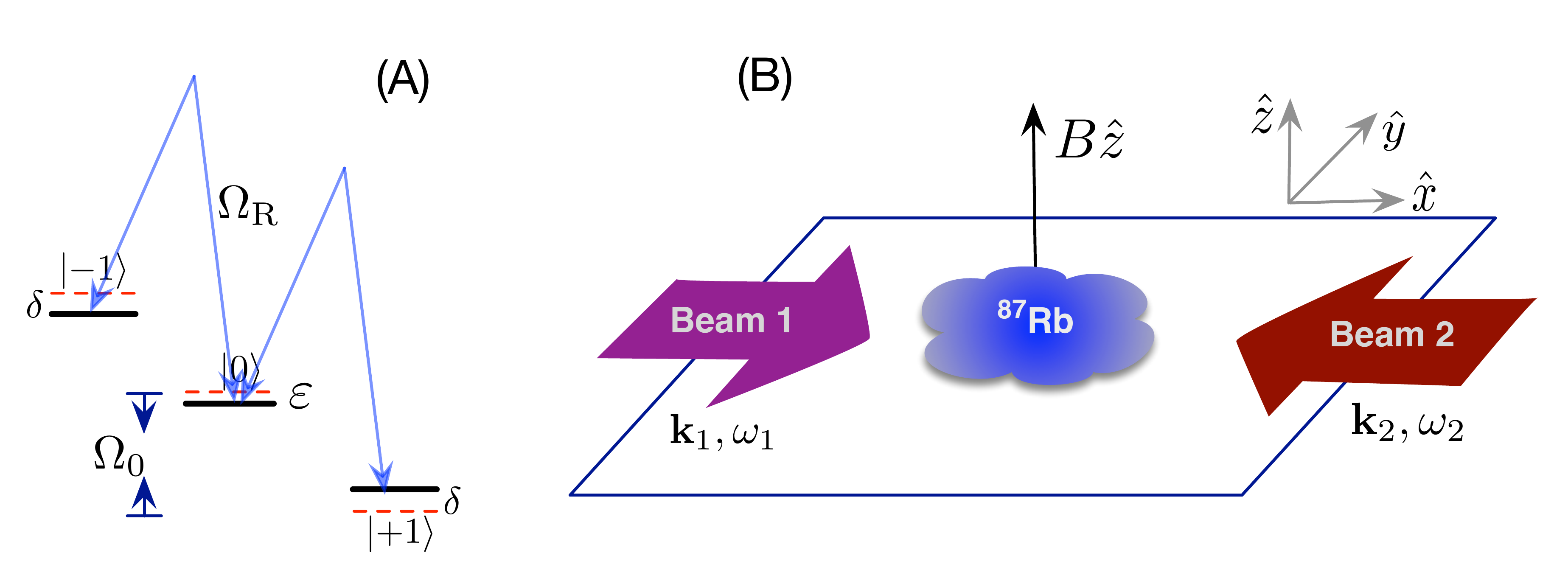}
\caption{Schematic setup of the NIST experiment. (A) The level diagram of $^{87}$Rb in its ground state $F=1$ manifold. The linear Zeeman splitting is given by $\hbar\O_0$. $\d=\o-\O_0$ is the detuning of the beams from the Raman resonance. $\O_{\rm R}$ is the two-photon Rabi frequency. $\ve$-term is the quadratic Zeeman effect which shifts the $F_z=0$ state downwards by amount $\hbar\ve$. (B) Two counter-propagating laser beams impinge on a cloud of $^{87}$Rb atoms along the $\pm\hat{x}$-axis. An external magnetic field is applied along the $\hat{z}$-direction. In the resulting adiabatic states, the atoms behave as charged particles in an external gauge field. Figure adapted from ref.\cite{Ho2011}}
\label{RamanSchematic}
\end{center}
\end{figure}

As a first step, we outline the Raman scheme for generating spin-orbit coupling in a uniform system~\cite{Higbie2002, Spielman2009}. As shown in Figure~\ref{RamanSchematic}, a pair of Raman beams with frequencies $\o_{1,2}$, wave vectors ${\bf k}_{1,2}$ and polarization $\hat{\l}_{1,2}$ are applied to the atomic $^{87}$Rb vapor. An external magnetic field ${\bf B}$ is applied along the $\hat{z}$-axis and sets the quantization axis of the hyperfine spin ${\bf F}$. The Raman lasers transfer momentum $2{\bf q}\equiv {\bf k}_1-{\bf k}_2$, which we take to be along the $\hat{x}$-direction, to the atom and at the same time flip its spin, depending on the polarizations of the two laser beams.

%--------------------------Figure--------------------------
%--------------------------Figure--------------------------

The single particle Hamiltonian has the form,
\be
H(t) = \frac{{\bf p}^2}{2m}  - \hbar\Omega_{0}F_{z} + \hbar \ve F_{z}^2 -\frac{\hbar \Omega_{R}}{2}\left[e^{i(2qx-\omega t)}(F_{x}+iF_{y}) + {\rm H.c.} \right]
\ee
where $\Omega_{0}$ is the Larmor frequency associated with the uniform external magnetic field along $\hat{z}$ direction and $m$ is the mass of the atom under consideration. The $\ve$-term arises from the quadratic Zeeman effect which shift the $F_z=0$ state downwards by amount $\hbar\ve$ (we have neglect a constant term $-\hbar\ve$ in the Hamiltonian). The Rabi frequency $\Omega_R$ describes the coupling between different spin states due to the laser fields and is referred to as the two-photon Rabi frequency.  The momentum and energy transfer between the atom and laser field are given by $2{\bf q}={\bf k}_1-{\bf k}_2\equiv 2q_x\hat{x}$ (say along $\hat{x}$-direction) and $\o=\o_1-\o_2$. The explicit time-dependence of $H(t)$ can be eliminated by performing a unitary transformation $U(t)=\exp(i\o t F_z)$, then $\widetilde{H}=U^\dag H(t) U$ is time-independent
\be
\widetilde{H}=\frac{{\bf p}^2}{2m}  +\exp(-i2qx F_z)\left[-\hbar(\Omega_{0}-\o)F_{z} + \hbar \ve F_{z}^2 -\hbar \Omega_{R} F_x\right]\exp(i2qx F_z)
\ee
This transformation describes the spins spiraling around the $\hat{z}$-axis with a period $\pi/q$. It is then possible to eliminate the spatial dependence $\exp(-i2qx F_z)$ by a similar unitary transformation on the Hamiltonian $\widetilde{H}$ and one then ends up with a spin-orbit coupled Hamiltonian of the form~\cite{Ho2011,Lin2012}
\be
H_{\rm so}=\frac{({\bf p}+qF_z\hat{x})^2}{2m}-\hbar(\Omega_{0}-\o)F_{z} + \hbar \ve F_{z}^2 -\hbar \Omega_{R} F_x.
\ee
In the ground state manifold of $^{87}$Rb atoms with $F=1$, depending on the choice of various parameters, this apparently simple Hamiltonian contains both the abelian synthetic gauge field and spin-orbit coupling as limiting cases. \\
(1) When $\O_{\rm R}\gg \ve, q^2/2m$ and $\o\approx \O_0$, the single lowest spin state is given by $F_x=-1$. In this case, one can project the Hamiltonian to this single state, and with a magnetic field gradient, one realizes the traditional $U(1)$ abelian gauge field. The formation of superfluid vortices has indeed been observed in experiment~\cite{Lin2009a} in this regime. \\
(2) When $\ve\gg \O_{\rm R}, q^2/2m$ and $\o\approx\O_0-\ve$ the two states with $F_z=0$ and $F_z=1$ are nearly degenerate and upon projecting the Hamiltonian to the space spanned by these two spin states, which we shall denote as (pseudo-spin) $\s$, a spin-orbit coupling is realized, with the Hamiltonian taking on the following form 
\be
\label{1Dlattice}
H_{\rm so}=\frac{(p_x+q\s_z)^2}{2m}+\frac{\d}{2}\s_z+\frac{\O_{\rm R}}{2}\s_x.
\ee
where $\d=\o-\O_0$ is the detuning from Raman resonance.

Now let us introduce an one-dimensional optical lattice with optical potential given by $V(x)=sE_{\rm R}\sin^2(Kx)$, where $K$ is the wave vector of the optical lattices and $E_{\rm R}\equiv \hbar^2 K^2/2m$ is the recoil energy and $s$ characterizes the depth of the potential. The single particle Hamiltonian becomes
\be
H_0=\frac{(p_x+q\s_z)^2}{2m}+\frac{\d}{2}\s_z+\frac{\O_{\rm R}}{2}\s_x+sE_{\rm R}\sin^2(Kx).
\label{solattice}
\ee
While the discussion below will be for one-dimensional case, it is straightforward to generalize it to higher dimensions.

It is helpful to construct an appropriate tight-binding (TB) model to describe this system in the limit of deep optical lattices $s\gg 1$. To do this, in principle one should solve for the band spectrum of the Hamiltonian in Eq.~(\ref{solattice}) and then construct the appropriate Wannier states. When the spin-orbit coupling is weak, there will be two nearly degenerate bands (corresponding to, roughly, the two spin components) with a large band gap $\D\propto \sqrt{s}$ to all other higher bands. As a result, we can concentrate on the lowest two nearly degenerate bands. Thus, we expect to find a TB model with two spin-resolved orbitals associated with every lattice site $W^{\rm I}_i({\bf r})$ and $W^{\rm II}_i({\bf r})$, each in general a superposition of spin-up and spin-down states. The appropriate hopping constant can be calculated as
\be
T_{ij}^{\tau\tau'}=\bra{W^{\tau}_i({\bf r})} H_0|W^{\tau'}_j({\bf r})\rangle; \quad \tau,\tau'={\rm I,II},
\ee
where $\tau$ and $\tau'$ label the two Wannier states. Writing this in terms of the original spin degree of freedom, one obtains the appropriate hopping Hamiltonian. While this work may be necessary for detailed quantitative comparisons between theory and experiments, we can reason from very general considerations to determine the general structure of the hopping model. In particular, by interpreting the spin-orbit coupling itself as a gauge field which enters the Hamiltonian in the minimal coupling form, we can invoke the Peierls' substitution to provide the correct lattice model. Thus, we can write the hopping matrix along $\hat{x}$-direction as~\cite{Grass2011,Cole2012,Radic2012}
\be
T_{\hat{x}}=t_{\hat{x}}\exp(-i\a A_{\hat{x}})=t_{\hat{x}}\exp\BK{i\frac{\pi q_x}{K}\s_z},
\ee
where we have used the fact that lattice constant $a=\pi/K$ and the gauge field along $\hat{x}$-direction is given by $A_{\hat{x}}=-q_x\s_z$. $t_{\hat{x}}$ is the hopping parameter in the absence of spin-orbit coupling. Since one can arrange the direction of momentum transfer $2{\bf q}$ to be different from the lattice direction, a similar term can be generated along $\hat{y}$-direction. Note that the coupling term would be $p_y\s_z$ instead of the isotropic Rashba form $p_x\s_y - p_y\s_x$. The strength of the spin-orbit coupling can be tuned by changing the value of $2{\bf q}$. In general, microscopic calculations with the correct Wannier states find that spin-orbit coupling modifies {\em both} $t_{\hat{x}}$ and the phases factors, but the general structure ($\s_z$ dependences) will be left un-modified as long as the two-band approximation remains appropriate. A careful comparison of the Peierls' substitution with a numerical calculation of the Wannier functions and hopping matrix elements was carried out in ref.~\cite{Radic2012}. We note in passing that in the experiments conducted so far, the momentum transfer is typically comparable to the lattice wavevector $K$ and are in the regime where Peierls' substitution begins to become quantitatively inaccurate.

\subsection{Laser assisted tunneling}
The physics of laser-assisted hopping can be illustrated most easily with a double-well potential~\cite{Kolovsky2011,Creffield2013}. Consider an atom with two internal states (i.e. spin-$1/2$) in a spin-independent double-well potential. For simplicity, let us further assume that around each minimum of the double-well, the oscillator frequencies are identical and are given by $\o_0$. The associated localized wave functions are given by $\varphi_{{\bf R}_i}({\bf r})$ and $\varphi_{{\bf R}_j}({\bf r})$, where ${\bf R}_{i}$ and ${\bf R}_{j}$ label the two wells along $\hat{x}$-direction with $a\equiv |{\bf R}_{i}-{\bf R}_{j}|$. Now, in the absence of a tilting potential, $\D=0$, the normal tunneling between these two sites is diagonal in spin space and is given by $J_0$; its value depends on the overlap of the two Wannier wave functions $\varphi_{{\bf R}_i}({\bf r})$ and $\varphi_{{\bf R}_j}({\bf r})$. When $\D\neq 0$, normal tunneling between the two sites ${\bf R}_i$ and ${\bf R}_j$ is suppressed due to the energy mismatch. To restore hopping, a pair of far-off resonant laser beams are applied with wave vectors ${\bf k}_1$ and  ${\bf k}_2$, frequencies $\o_1$ and $\o_2$. This induces a coupling term of the following form,
\be
V_{\rm laser}=\hbar\O_{\rm R}\BK{\exp(i2{\bf q}\cdot{\bf r}-i\o t)\hat{\mathcal{S}}+\exp(-i2{\bf q}\cdot{\bf r}+i\o t)\hat{\mathcal{S}}^\dagger},
\ee
where $\O_{\rm R}$ is the two-photon Rabi frequency and recall that $2{\bf q}={\bf k}_1-{\bf k}_2\equiv 2q_x\hat{x}$ and $\o=\o_1-\o_2$ are the momentum and energy transfer between the atom and the laser field. $\hat{\mathcal{S}}$ describes the action of the two laser beams on the spin states of the atom and depends on the polarization of the two laser beams. If $\hat{\mathcal{S}}$ is diagonal in spin space, the result is two decoupled optical lattices for each spin component. Each copy can feature a complex hopping amplitude, which in general can give rise to abelian gauge fields. On the other hand, if $\hat{\mathcal{S}}$ is non-diagonal in spin space, this may be used to realize a non-abelian gauge field in an optical lattice, of which spin-orbit coupling is a special case. It is worthwhile to point out that a non-digonal $\hat{\mathcal{S}}$ relies on the internal atomic spin-orbit coupling, and this can lead to siginificant spontaneous emission in alkali atoms~\cite{Kennedy2013}.

%-----------------------figure--------------------------------------------------
\begin{figure}[ht]
\begin{center}
\includegraphics[width=0.8 \textwidth]{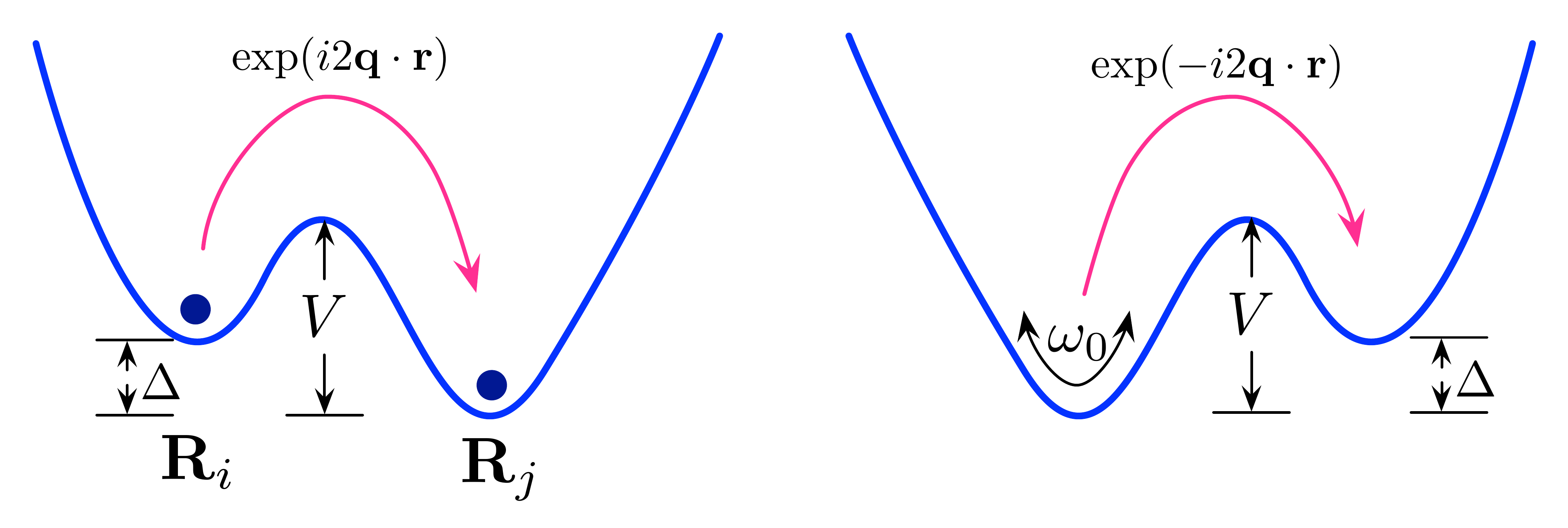}
\caption{Laser assisted hopping between two wells with energy offset $\Delta$ and a potential barrier $V$. The harmonic frequency at the bottom of the well is $\o_0$. When an atom from the higher energy well at ${\bf R}_i$ tunnels to the lower energy well at ${\bf R}_j$, an energy of $\D$ must be absorbed by the radiation field with a spatially-varying phase factor $\exp(i2{\bf q}\cdot{\bf r})$. The hopping from ${\bf R}_j$ to ${\bf R}_i$, on the other hand, will be associated with a phase factor $\exp(-i2{\bf q}\cdot{\bf r})$.}
\label{fig:lah}
\end{center}
\end{figure}
%-----------------------figure--------------------------------------------------

Resonant hopping can be restored when $\hbar\o=\D$. In the dressed atom picture, this means that an atom initially residing at the higher potential ${\bf R}_i$ with $n_1$-photon in mode $(\o_1,{\bf k}_1)$ and $n_2$-photon in mode $(\o_2,{\bf k}_2)$, is resonant with a state in which the atom is at the lower potential ${\bf R}_j$ with $(n_1+1)$-photon in mode $(\o_1,{\bf k}_1)$ and $(n_2-1)$-photon in mode $(\o_2,{\bf k}_2)$. Crucially, the spatial phase associated with hopping from ${\bf R}_i$ to ${\bf R}_j$ is given by $\exp(i2{\bf q}\cdot{\bf r})$, while that from ${\bf R}_j$ to ${\bf R}_i$ is given by $\exp(-i2{\bf q}\cdot{\bf r})$. Thus the laser beams imprint a complex Peierls' phase during the hopping process. It can be shown that the effective hopping amplitude as modified by the laser beams is given by
\be
J_{\rm eff}\exp(-i{\bf q}\cdot({\bf R}_i+{\bf R}_j)),
\ee
where $J_{\rm eff}=J_0\mathcal{J}_1(\kappa)$, where $\kappa={4\O}\sin(q_xa)/{\D}$, depending on the details of the laser arrangements and the distance between the neighboring sites. $\mathcal{J}_1(\kappa)$ is the first order Bessel function.  

The idea of Raman assisted tunelling in optical lattices is proposed in ref.~\cite{Jaksch2003} and later extended in ref.~\cite{Gerbier2010}. Within the Raman scheme, two types of magnetic flux patterns have been realized to date. In the first experiment from Munich, a staggered magnetic flux along one-direction was realized with a superlattice potential~\cite{Aidelsburger2011,Aidelsburger2013a}. The magnetic flux per plaquette can be tuned easily by changing the angle between the two laser beams. In later experiments from both Munich~\cite{Aidelsburger2013b} and MIT~\cite{Miyake2013}, uniform flux is realized with a linear potential generated by the magnetic field gradient. Extension of the Raman scheme to create spin-orbit coupling is discussed in ref.~\cite{Kennedy2013}.

\subsection{Periodically driven lattice}
Another way of generating artificial magnetic fields and spin-orbit coupling is to use periodically driven systems~\cite{Eckardt2010,Goldman2014}. In this scheme, a time-periodic Hamiltonian is considered $\hat{H}(t)=\hat{H}(t+T)$. The time evolution of the system is described by the evolution operator $\hat{U}(t)=\mathcal{T}\exp[-i\int_0^t\hat{\mathcal{H}}(t')dt']$. Because of the periodicity of the problem, one looks for the evolution operator over a period $\hat{U}(T)$ and defines an effective Hamiltonian
\be
\hat{U}(T)=\mathcal{T}\exp[-i\int_0^T\hat{\mathcal{H}}(t')dt']\equiv \exp[-i T H_{\rm eff}].
\ee
The form of $H_{\rm eff}$ can be very complicated and no closed form exists in general. However, provided that the modulation frequency $\o=2\pi/T$ is large compare with typical energy scales in the problem and the modulation amplitude is small, it is possible to develop a formal expansion in $1/\o$. Let us write $\hat{H}(t)$ as a Fourier series
\be
\hat{H}(t)=\sum_{n=-\infty}^\infty\hat{H}_n\exp[in\o t],
\ee
then the effective Hamiltonian up to first order in $1/\o$ is given by~\cite{Goldman2014,Jotzu2014}
\begin{align}
\hat{H}_{\rm eff}=\hat{H}_0+\frac{1}{\o}\sum_{n=1}^{\infty}\frac{1}{n}[\hat{H}_n,\hat{H}_{-n}].
\end{align}
We note that the zeroth order term is just the time average of the Hamiltonian over a period $T$. Typically, this modulation is applied in combination with an optical lattice (``shaking lattice") and the time dependence enters into the Hamiltonian by coupling to a term of the form
\be
\sum_{\s\s'{\bf r}}\hat{H}_{\rm mod}({\bf r},t)a_{\s{\bf r}}^\dagger a_{\s'{\bf r}},
\ee
where the modulation coupling $\hat{H}_{\rm mod}({\bf r},t)$ can be spatially varying and, furthermore, can be a matrix in spin space. A few examples that have been realized in recent experiments are given in Table~\ref{shaking_exp}. Extensions of the shaking scheme to generate spin-orbit coupling are discussed in Refs.~\cite{Struck2014,Goldman2014}.

%--------------------------Table--------------------------
\begin{table}[h]
\def\arraystretch{1.5}
\setlength{\tabcolsep}{0.5 em}
\begin{center}
  \caption{\label{shaking_exp}Different lattices and physical models realized so far with the periodic driven lattice technique. In the table, $\o$ is the driving frequency and $T$ is the period of the driving. $\hat{e}_{1,2}$ are two orthonormal vectors in the $xy$-plane. $\d{\bf k}$ is the momentum transfer from the Raman beams. $\O_{\rm R}$ is the Rabi frequency. $F$ and $F_{1,2}$ are the amplitude of the driving.}
    \begin{tabular}{ | c | c | c | c |}
    \hline	
      Group & underlying lattice  & driven term ($\hat{H}_{\rm mod}({\bf r},t)={\bf F}(t)\cdot{\bf r}$)  & physical models\\[3pt]  \hline	
    Hamburg & 1D lattice & ${\bf F}(t)=F\sin(\omega t), t<T_1; ~{\bf F}(t)=0, T_1<t<T$ & Peierls phase~\cite{Struck2012} \\[3pt]  \hline	
    Hamburg & triangular lattice & {${\bf F}(t)=F_1\cos(\omega t)\hat{e}_1+F_2\sin(\o t)\hat{e}_2$} & frustrated spin model~\cite{Struck2011}\\[3pt]  \hline
    Hamburg & triangular lattice & ${\bf F}(t)=F_1\cos(\omega t)\hat{e}_1+F_2[\sin(\o t)+\d\sin(2\o t)]\hat{e}_2$ & Ising-XY spin model~\cite{Struck2013} \\[3pt]  \hline
    Munich & superlattice & $\O_{\rm R} \sin(\d{\bf k}\cdot{\bf r}-\o t)$ & staggered flux~\cite{Aidelsburger2011}\\[3pt]  \hline
    \pbox{2cm}{Munich\\MIT}	& \pbox{2.4 cm}{optical lattice+ \\linear potential} &$\O_{\rm R} \sin(\d{\bf k}\cdot{\bf r}-\o t)$ & uniform flux~\cite{Aidelsburger2013b, Miyake2013}\\[3pt]  \hline
      Chicago & 1D lattice & $U_0\sin^2(k(x-x_0(t)))$ & ferromagnetic domain~\cite{Parker2013}\\[3pt]  \hline
     ETH	& honeycomb lattice & ${\bf F}(t)=F[\cos(\omega t)\hat{e}_1+\cos(\o t-\varphi)\hat{e}_2]$ & Haldane model~\cite{Jotzu2014}\\[3pt] 
    \hline
    \end{tabular}
\end{center}
\def\arraystretch{1}
\end{table}
%--------------------------Table--------------------------

\subsection{Zeeman lattice}
The concept of a ``Zeeman lattice" was introduced in ref.~\cite{Garcia2012}, where a combination of Raman and radio-frequency laser beams produce an effective magnetic field that varies periodically both in its {\em magnitude} and {\em direction}. This is related to the more general idea of optical flux lattices, introduced in ref.~\cite{Cooper2011}. As before, consider $^{87}$Rb atoms with the $F=1$ ground state split due to a Zeeman field, as shown in Fig.~\ref{RamanSchematic}. In addition to the Raman beams, one add an extra radio-frequency (RF) beam which drive direct transitions between hyperfine-Zeeman levels. The coupling strength and frequency of the RF field is given by $\O_{\rm rf}$ and $\o_{\rm rf}$, respectively. One also defines the detuning as $\d=\o_{\rm rf}-\O_0$. It is necessary that the energy transfer from the Raman beams be the same as the rf beam, in order that one can go to a common rotating frame. As a result, the effective Hamiltonian is given by
\be
\hat{H}_{\rm rf+Raman}=\frac{p^2}{2m}\hat{I}+\boldsymbol{\O}(x)\cdot\hat{\bf F}+\hat{H}_Q,
\ee
where $\hat{I}$ is a $3\times 3$ identity matrix. $H_Q=-\varepsilon(\hat{I}-F_z^2)$ is the quadratic Zeeman shift. $\hat{\bf F}=(F_x,F_y,F_z)$ and the effective magnetic field is given by~\cite{Garcia2012}
\be
\boldsymbol{\O}(x)=\frac{1}{\sqrt{2}}(\O_{\rm rf}+\O_{\rm R}\cos(2q x),-\O_{\rm R}\sin(2q x),\sqrt{2}\d),
\ee
where $2q$ is the momentum transfer from the Raman beams. Without the rf-field, one finds a Zeeman field whose direction is rotating in the $xy$-plane, but the amplitude stays the same. This realizes the standard spin-orbit coupling in the uniform system. With additional rf-field, the magnitude of the Zeeman coupling, $|\boldsymbol{\O}(x)|$, is changing periodically and provides a one-dimensional ``Zeeman lattice". When an atom hops from one minimum of the lattice to its nearest neighbors, the effective magnetic field winds in the Bloch sphere and generate a geometric Berry phase~\cite{Garcia2012}. An experiment using $^{87}$Rb has measured the Peierls' phase generated and also the effective mass close to the band minimum~\cite{Garcia2012}. The spin resolved band structure in a ``Zeeman lattice" has been mapped out using fermionic $^{6}$Li with a novel spin injection spectroscopy technique in ref.~\cite{Cheuk2012}.

\section{Basic Theoretical Model}

\label{sec:btm}
Having now discussed several experimentally viable routes to implementing spin-orbit coupling in an optical lattice, we next turn our attention to new many-body physics
which results from the interplay of spin-orbit coupling, lattice environment, and interactions. We do not tie ourselves to any specific experimental realization, assuming that
specific model Hamiltonians we consider can be realized using schemes discussed above or suitable variants. The appropriate degrees of freedom will be boson or fermion operators associated with Wannier states localized near the lattice sites, which also carry a (typically two-component) pseudospin degree of freedom. For a sufficiently deep lattice potential that the occupation of excited bands can be neglected, the resulting system is well-described by a tight-binding hamiltonian
\be
H_0=-t\sum_{\lr{ij}}a_{i\s}^\dag\mathcal{R}_{ij}^{\s\s'}a_{j\s'} + \mbox{h.c.},
\ee
where $t$ sets an overall hopping amplitude (and natural energy scale), while $\mathcal{R}_{ij}^{\s\s'}$ characterizes the hopping of an atom with spin $\s'$ from site $j$ to the spin $\s$ state on a neighboring site $i$. Here we have dropped the Zeeman terms associated with Rabi frequency and detuning in order to define a minimal model to investigate the interplay between spin-orbit coupling and strong interactions. Because of the lattice-translation invariance of this hamiltonian, it can be Fourier transformed into ${\bf k}$-space, and the Hamiltonian can be generically written as
\be
H_0=\sum\psi_{\bf k}^\dag \mathcal{H}({\bf k})\psi_{\bf k}
\ee
where $\mathcal{H}({\bf k})=d_0({\bf k})+{\bf d}({\bf k})\cdot\boldsymbol{\s}$, is written in terms of its expansion in the Pauli matrices $\boldsymbol{\s}=(\s_x,\s_y,\s_z)$. The topological properties of any Hamiltonian expressed in this way can be obtained from a straightforward computation of the Berry curvature (for two spatial dimensions)
\be
F_{ij}=\frac{1}{2}\e_{abc}\hat{d}_a\partial_i \hat{d}_b \partial_j \hat{d}_c 
\ee
where $\hat{d}_a=d_a/|{\bf d}|$. The Chern number can then be computed by integrating the Berry curvature over the occupied states\cite{Bernevig2013}.

In the following, we shall explicitly work through two important examples: (1) Anisotropic spin-orbit coupling (e.g., $p_x\s_y$) in a 1D optical lattice and (2) isotropic Rashba spin-orbit coupling in 2D optical lattices. The first case has been realized in an experiment which explores the stability of the spin-orbit coupled condensate in a moving 1D optical lattice\cite{Hamner2014}. The second case has not yet been realized, but is the subject of a substantial experimental effort.

\subsection{Band structure}

\label{sec:bs}

(1) \emph{One-dimensional spin-orbit coupling in a one dimensional lattice.} This is the case with the present implementations of spin-orbit coupling, where only one component of the momentum (say $k_x$) is coupled nontrivially to the spin. The hopping matrix is then given by
\be
\lb{hopx}
\mathcal{R}^{\hat{x}}_{ij}=\cos\a\pm i\sin\a\s_y
\ee
where $\pm$ refers to hopping along the $+\hat{x}$ or $-\hat{x}$ directions, and reveals the explicitly broken spatial inversion symmetry. The single-particle spectrum is given by two bands with $\e_\pm(k)=-2t\cos(k\pm \a)$. There are two degenerate minima at $k=\pm \a$ and the associated spinor wave functions are $\chi_{\pm,\a}(x)=\frac{1}{\sqrt{2}} \exp(\pm i\a x)(1, \pm i)^T$. These two states form a pair, $\chi_{+,\a}(x) = -i \sigma_x \chi_{-,\a}(-x)$, related to each other by inversion followed by a spin rotation by $\pi$ about the $\sigma_x$
axis. At zero momentum ($k=0$) the two bands have a level crossing, leading to a doublet protected by time-reversal symmetry. This Kramers degeneracy is broken in the presence of a symmetry-breaking Zeeman term, as appears in present experiments with Raman-induced spin-orbit coupling. Near this avoided crossing, the energy spectrum is described by a one-dimensional massive Dirac equation.

%-----------------------figure--------------------------------------------------
\begin{figure}[h]
\begin{center}

%\begin{subfigure}[c]{0.5\textwidth}
\includegraphics[width=0.5\textwidth]{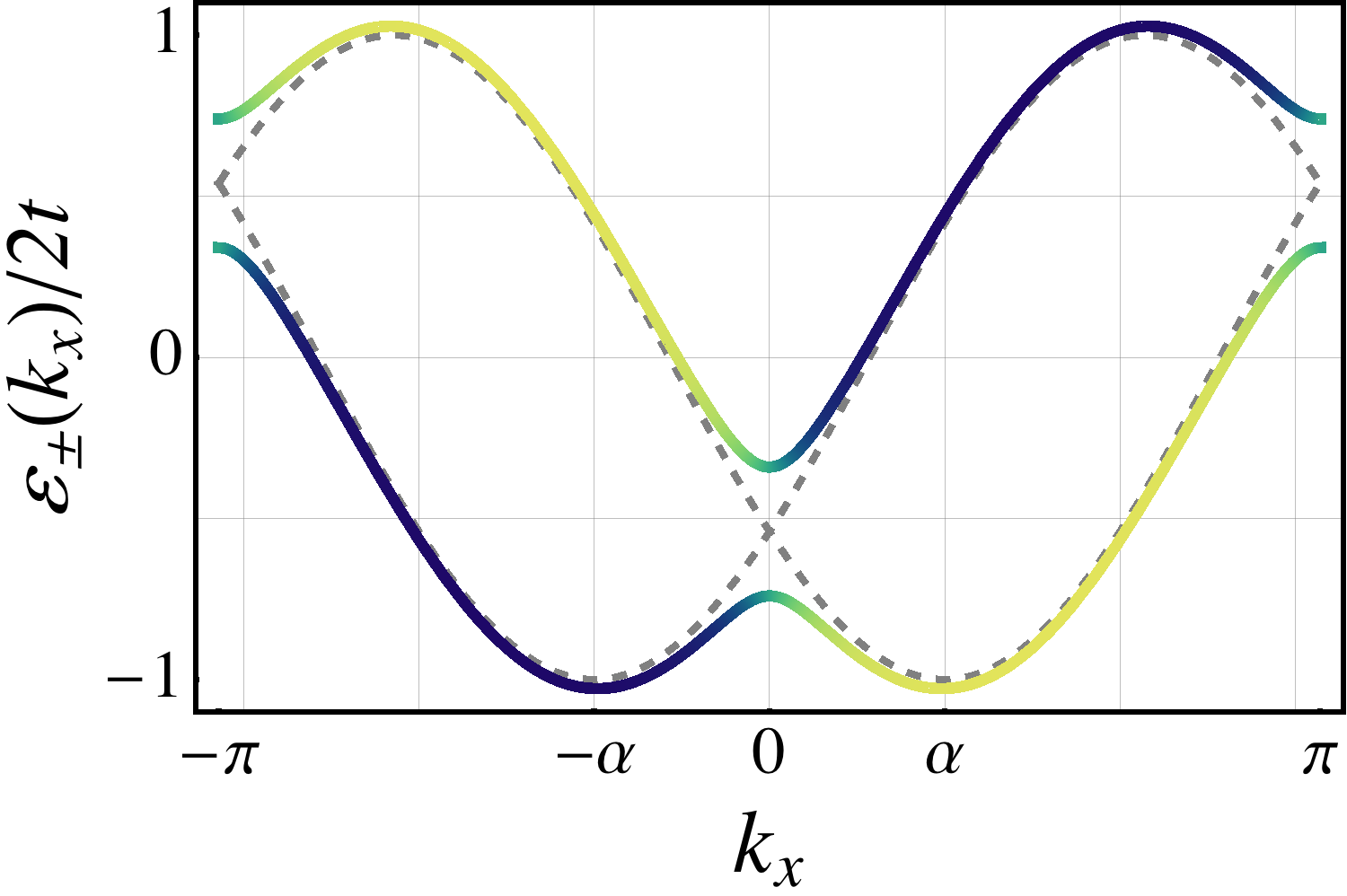}
%\end{subfigure}
\hspace{1cm}
%\begin{subfigure}[c]{0.35\textwidth}
\includegraphics[width=0.35\textwidth]{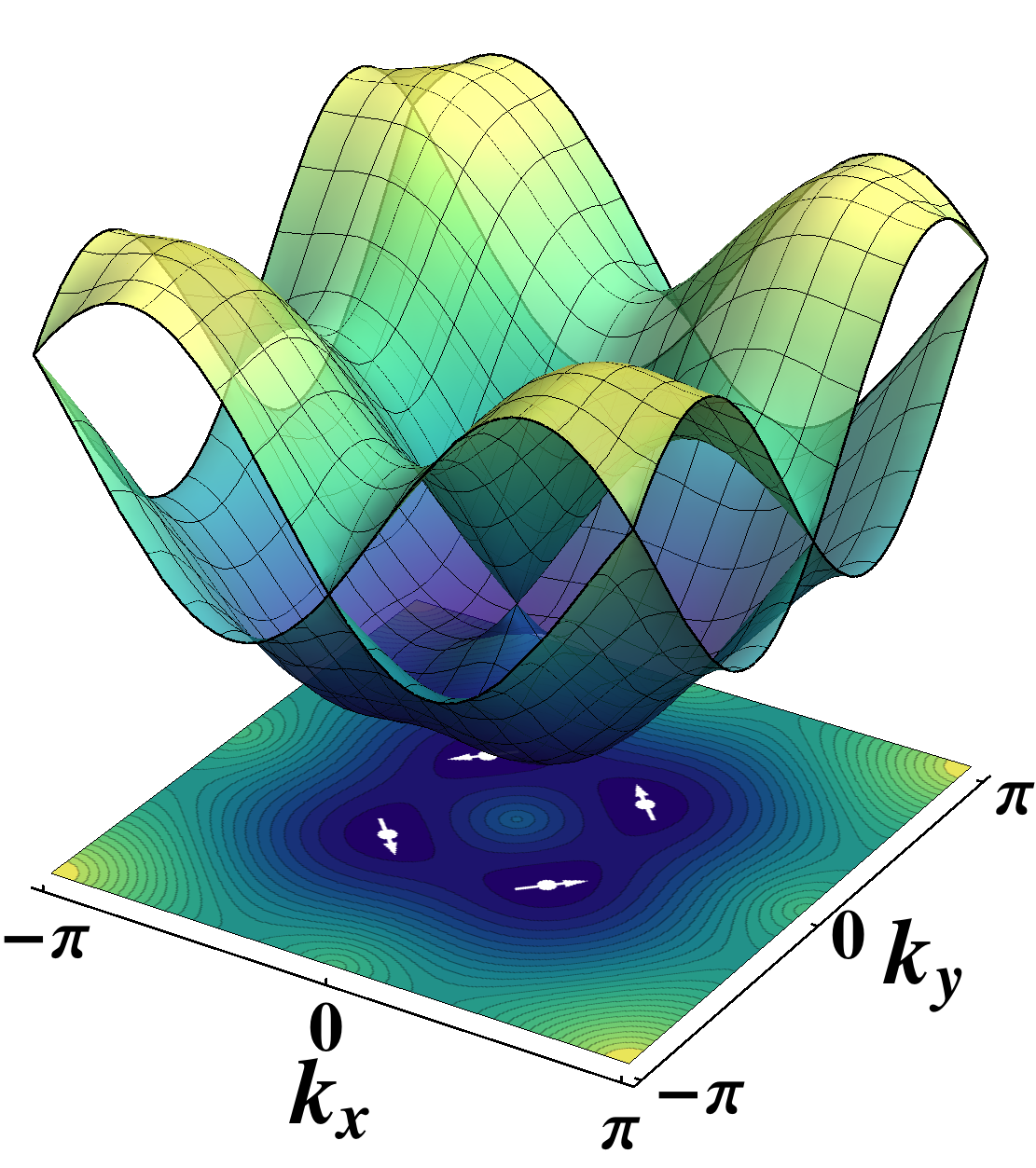}
%\end{subfigure}

\caption{(Left) Energy eigenvalues $\e_\pm(k_x)=-2t\cos(k_x \pm \alpha)$ resulting from the one-dimensional spin-orbit coupling Eq.~(\ref{hopx}). Additionally introducing a small Zeeman coupling splits the two bands, generating avoided crossings at $k_x=0,\pi$. The color corresponds to the $y$-component of spin $\lr{\sigma_y}$ in that state. Dark blue corresponds to states where the spin is locked to the $-y$ direction, yellow to $+y$. (Right) Energy bands arising from Rashba spin-orbit coupling in a square lattice. Here, the color simply tracks the energy. The momenta corresponding to the four lowest-energy states are marked with white dots and arrows which represent the spin wavefunction associated with those states. The right-side figure is adapted from ref.\cite{Cole2012}}
\label{fig:1d_2d_spectrum}
\end{center}
\end{figure}
%-----------------------figure--------------------------------------------------

(2) \emph{Rashba spin-orbit coupling in a two-dimensional square lattice.} In this case, in addition to the hopping along the $\hat{x}$-direction, Eqn.(\ref{hopx}), there appears an additional contribution from hopping along the $\hat{y}$ direction, given by the matrix
\be
\lb{hopy}
\mathcal{R}^{\hat{y}}_{ij}=\cos\b\pm i\sin\b\s_x,
\ee
which links motion along the $\hat{y}$ direction to the spin projection along $\hat{x}$. In general, $\a$ and $\b$ can be different, resulting in an arbitrary linear combination of the linear Rashba and linear Dresselhaus spin-orbit couplings.

Using the expansion in Pauli matrices, the tight-binding Hamiltonian can be characterized by the $d$-vector
\begin{align}
d_0({\bf k}) &= -2t \left( \cos \a \cos k_x + \cos \b \cos k_y \right) , \\
d_x({\bf k}) &= -2t \sin \b \sin k_y,\\
d_y({\bf k}) &= -2t \sin \a \sin k_x,\\
d_z({\bf k}) &=0
\end{align}
The energy spectrum is easily evaluated, as $\e_{\pm}(\vec{k}) = d_0(\vec{k}) \pm |\vec{d}(\vec{k})|$, while the spin eigenstates with $d_z = 0$ are
\be
\chi_{\pm}(\vec{k}) = \left( \begin{array}{c} 1 \\  \mp i e^{i\varphi_{\vec{k}}} \end{array} \right)
,\quad \varphi_{\vec{k}} = \arctan \left( d_x/d_y \right)
\ee
which has singularities whenever $d_x=d_y=0$. In the first Brillouin zone, this occurs at the four time-reversal invariant momenta $(k_x,k_y) \in \{(0,0),(0,\pi),(\pi,0),(\pi,\pi)\}$, and signals the locations of Dirac points. The circulation of the spin wavefunction is counter-clockwise for contours that enclose $\vec{k}=(0,0)$ or $(\pi,\pi)$ (in the usual counter-clockwise sense), and clockwise for those that surround $\vec{k}=(0,\pi)$ or $(\pi,0)$. These two species of Dirac points are therefore topologically distinct from one another, with winding numbers $\pm 1$.

\subsection{Strong interaction physics}
\label{sec:sip}
One of the primary reasons for interest in optical lattices is that they provide a route to tunable strong interactions between particles. Working in the regime where the interaction scale $U \ll \D$, with $\D$ the gap to the lowest excited band, it is possible to write the interaction contribution to the Hamiltonian as
\be
H_{\rm int}= \sum_{i} \left( \frac{U}{2} \sum_{\s} \left[ n_{i\s} (n_{i\s}-1) \right] + U' n_{i\ua}n_{i\da} \right) +\ldots,
\ee
where $\ldots$ accounts for (typically negligible) further neighbors interactions. $U$ and $U'\equiv\l U$ describe the intra- and inter-species interactions. Considering only the on-site interaction, the full model Hamiltonian has the form
\be
\lb{sobhm}
H=-t\sum_{\lr{ij}}a_{i\s}^\dag\mathcal{R}_{ij}^{\s\s'}a_{j\s'} +
\sum_{i} \left( \frac{U}{2} \sum_{\s} \left[ n_{i\s} (n_{i\s}-1) \right] + U' n_{i\ua}n_{i\da} \right) .
\ee
We shall refer to this as spin-orbit coupled Bose-Hubbard model (SOBHM). In the following, we discuss a few special features that occur in the combined presence of spin-orbit coupling and strong inter-particle interaction.

\subsection{General Discussions}
There are already several new physical effects associated with spin-orbit coupling that have been demonstrated in recent experiments. In the case of bosons, for example, depending on the parameters, the single particle ground state can be degenerate and the Bose condensate can feature novel density and spin density patterns. A further consequence of spin-orbit coupling is the lack of Galilean invariance, as demonstrated in the recent moving optical lattice experiment. Furthermore, in a harmonic trap, spin-momentum locking provides a way to couple the dipole oscillation to magnetic oscillations. In this review, we focus on interesting new features that are brought about by spin-orbit coupling in connection with strong interaction effects.

{\bf 1. Exotic magnetic structures in Mott insulators}. A natural question regarding SOBHM is the magnetic phases deep in the Mott insulating regime. This has been addressed in several works~\cite{Cole2012,Radic2012,Cai2012,gong_arXiv_2012}. For the standard Bose-Hubbard model with spinless bosons, the Mott insulating state is a featureless Mott insulator with a charge gap and zero compressibility. However, introducing a spin degree of freedom as well as the spin-orbit coupling present in $H_0$ allows for the realization of a rich class of magnetically ordered Mott insulators, similar to the generic form introduced by Moriya~\cite{Moriya1960} for electronic Mott insulators with spin-orbit coupling. To ${\cal O}(t^2/U)$, this hamiltonian is
\be
H_{\rm mag} = \sum_{i,\mu} \left[
J \vec{S}_i \cdot \vec{S}_{i+\mu} +
\vec{D}_{\mu} \cdot \left( \vec{S}_i \times \vec{S}_{i+\mu} \right) +
\sum_{a,b} S^a_i  \Gamma^{ab}_{\mu} S^b_{i+\mu} \right].
\label{eq:exchange}
\ee
where $\mu$ represents the spatial direction of a bond of the lattice. For concreteness, in later sections we will take a 1D chain along $\hat{x}$ and a 2D square lattice with lattice vectors $\hat{x}$ and $\hat{y}$. A detailed derivation of the above hamiltonian and the appropriate coefficients for these cases is provided in Appendix~\ref{app:strong_coupling}.

The natural energy scale here is given by $\mathcal{J} = \frac{4t^2}{\l U}$. In terms of this scale, we can write the exchange constant $J= -\mathcal{J} \cos 2\a$ which accompanies the spin-isotropic Heisenberg interaction; this is clearly ferromagnetic in the limit of vanishing spin-orbit coupling, as it must be for bosons. The remaining terms account for anisotropies in spin space that are generated either by the explicitly spin-anisotropic interactions of the bosons ($U' \neq U$) or by the explicit coupling of spin to orbital motion ($\a \neq 0$). The vectors $\vec{D}_{x}=- \left( \mathcal{J} \l \sin2\a \right)\hat{y}$ and $\vec{D}_{y}=-\left( \mathcal{J}  \l \sin2\a \right)\hat{x} $ arise purely from SOC, and characterize the antisymmetric Dzyaloshinsky-Moriya interaction \cite{Dzyaloshinsky1958,Moriya1960} which generically leads to long-wavelength magnetic spirals in solid-state materials. The SOC also generates symmetric anisotropic interactions of a ``compass model" type, $\Gamma^{xx}_{y}=\Gamma^{yy}_{x}= -\mathcal{J}(1-\cos 2\a)$, while an out-of-plane anisotropy  $\Gamma^{zz}_{x}=\Gamma^{zz}_{y}= -2\mathcal{J}(\l-1)$ arises from the original spin-anisotropy in the interactions (all other components of the tensor $\Gamma^{ab}_{\mu}$ are zero). Compass model interactions have become a major research topic of late for their role in Kitaev's exactly-solvable ``honeycomb model" of a spin liquid~\cite{Kitaev2006}, which itself might describe the magnetism of certain transition metal oxides with strong spin-orbit coupling~\cite{Jackeli2009}.

In the combined presence of these terms, the magnetic hamiltonian is generically frustrated and can support a wide variety of complex magnetic structures, including spiral and skyrmion states. This is discussed further in the following sections.

{\bf 2. The Mott transition from a non-uniform superfluid state}. In the standard BHM with only on-site interactions, both the superfluid state and the Mott state have uniform density, and the Mott-superfluid transition is accompanied by the breaking of $U(1)$ symmetry. On the other hand, when one considers the SOBHM, the superfluid state can exhibit spin density wave, while the spin density of the magnetically ordered Mott states is also generally inhomogeneous. Thus, apart from the usual broken $U(1)$ symmetry of the Mott-superfluid transition, there are order parameters associated with broken lattice translation symmetry in the Mott and superfluid states. As we shall discuss later, the magnetic structure can also persist across the Mott-superfluid transition.

These considerations imply that some generalizations to the standard treatment of BHM need to be made when considering the quantum phase transitions in a SOBHM. At the most na\"{i}ve level, since the superfluid state is no longer uniform, it is not possible to use the uniform Gutzwiller approximation, as is often done to describe mean-field properties of the BHM. It is at least necessarily to perform the mean-field analysis in a finite cluster. This has been explored in some detail by~\cite{Cole2012,mandal_PRB_2012,qian_arXiv_2013}. Beyond this level of mean-field theory, recently a bosonic variant of the \emph{dynamical} mean-field theory (BDMFT) has been applied~\cite{he_arXiv_2014}. Other, more exact, numerical methods are available in one dimension. The presence of an extra spin-density wave order parameter implies that the effective field theory of the Mott-superfluid transition could be quite different from the standard one and further research in this direction is worthwhile.

{\bf 3. Topological states in the SOBHM}. In the simplest case of a square lattice with Rashba spin-orbit coupling, the resulting single-particle band structure is topologically trivial, in the sense that the bands have zero Chern number. However even (or especially) in the absence of a nonzero single-particle Chern number, it is very interesting to ask if interaction effects can lead to nontrivial topological properties. The most natural place to look for such non-trivial topological properties is the Mott insulating state, where the single particle excitation spectrum is gapped. 

Having pointed out several new features that are likely to be encountered with spin-orbit coupling in the presence of Hubbard-type interactions, we proceed in the following sections with a few illustrative examples.

%This question has recently been answered in the affirmative by Wong and Duine, although they have only explored a few parameter regimes.

\subsubsection{One-dimensional lattice with spin-orbit coupling}
Let us consider first the case of a one-dimensional SOBHM, for which more exact treatment using density matrix renormalization group is possible. For simplicity, we shall neglect altogether the Zeeman terms and concentrate on the interplay between spin-orbit coupling and interactions. The Hamiltonian is given by Eqn.(\ref{sobhm}) with the hopping matrix $\mathcal{R}$ given by Eqn.(\ref{hopx}). Using intra species interaction $U$ to set the energy scale, we have $\a$, $t/U$ and $\l\equiv U'/U$ as three independent dimensionless parameters. 

In the strong coupling limit $t/U\ll 1$ with unit filling, the system enters Mott state with one boson per site and one obtains the effective magnetic Hamiltonian by the standard perturbation theory. In order to put the magnetic Hamiltonian in the standard form, we rotate the spin around $\hat{x}$-axis by $\pi/2$, such that the Dzyaloshinsky-Moriya vector is along $\hat{z}$-axis,
\begin{align}\label{magH1D}
H_{\rm mag}=-\frac{4t^2}{U}\sum_{\lr{ij}}\Big[\frac{\cos(2\a)}{\l}s^x_is^x_j+\frac{\cos(2\a)}{\l}(2\l-1)s^y_is^y_j+\frac{1}{\l}s^z_is^z_j+\sin(2\a)(s^x_is^y_j-s^y_is^x_j)\Big].
\end{align}
We note the following features: (1) The overall exchange energy scale is given by $t^2/U$ as usual, but the sign can be tuned by changing $\a$ and can be both ferromagnetic or antiferromagnetic; (2) There appears the Dzyaloshinsky-Moriya term, in addition to the standard Heisenberg coupling, with Dzyaloshinsky-Moriya vector given by ${\bf D}=\sin(2\a)\hat{z}$. $H_{\rm mag}$ cannot be solved exactly for general $\a$ and $\l$, but in various limits, it can be reduced to known models or exactly solvable~\cite{Zhao2014a,Zhao2014b,Xu2014,Piraud2014,Peotta2014}. The full phase diagram of $H_{\rm mag}$ is given in Fig.\ref{1dmag}. In the following, we consider a few special cases of $H_{\rm mag}$.\\
(1) For $SU(2)$ invariant interaction, {\it i.e.} $\l=1$, $H_{\rm mag}$ reduces to
\begin{align}
H_{\rm mag} = -\cos(2\a)\frac{4t^2}{U}\sum_{\lr{ij}}\Big[s^x_is^x_j+s^y_is^y_j  + \frac{1}{\cos(2\a)}s^z_is^z_j + \tan(2\a)(s^x_is^y_j-s^y_is^x_j)\Big].
\end{align}
This Hamiltonian can be transformed to an isotropic Heisenberg model if we make the following transformation~\cite{Perk1976}. At each site, the spin is rotated around $\hat{z}$-axis by an angle $\th_i$,  $\tilde{s}_i^+ \equiv\exp(-i\th_i s_z)s_i^+ \exp(i\th_i s_z) = \exp(-i\th_i)s_i^+$, where $s_i^+=s_x+is_y$ is the spin raising operator, while $\tilde{s}_i^z=s_i^z$. Choosing $\th_{i+1}-\th_i=-2\a$, the Hamiltonian \ref{magH1D} reduces to an isotropic ferromagnetic Heisenberg model in terms of the $\tilde{s}_i$ spins for any $\a$. That is, $H_{\rm mag}=-\frac{4t^2}{U}\sum_{ij}[\tilde{s}^x_i\tilde{s}^x_j+\tilde{s}^y_i\tilde{s}^y_j+\tilde{s}^z_i\tilde{s}^z_j]$. The exact ground state is a ferromagnet and the elementary excitations are spin waves with quadratic dispersion. In terms of the original spin $s$, this corresponds to an exact spiral ground state with wave vector $2\a$ along the chain.\\
(2) When $\a=\frac{\pi}{4}$, $H_{\rm mag}=-\frac{4t^2}{U}\sum_{ij}[\frac{1}{\l}s^z_is^z_j+(s^x_is^y_j-s^y_is^x_j)]$. This is a one-dimensional Ising model with DM interactions and has been studied in the literature~\cite{Jafari2008}. It has two phases: for $\l>1$, the DM term dominates and the system is in a chiral phase in which the spin spirals around the $\hat{z}$-axis along the chain. We refer to this as the chiral $xy$-magnet since the interactions will kill long range spin order but preserve the chirality. For $\l<1$, the ferromagnetic term dominates and the system is in a ferromagnetic state, pointing along the $\hat{z}$ direction. The ferromagnet has a twofold ground state degeneracy.\\
(3) A particularly interesting limit corresponds to taking $\l\to \infty$. In this case, the spin model is given by
\be
H_{\rm mag}=-\frac{4t^2}{U}\sum_{\lr{ij}}\left[2\cos(2\a)s^y_is^y_j+\sin(2\a)(s^x_is^y_j-s^y_is^x_j)\right].
\ee
This can be solved by the standard Jordan-Wigner transformation. The final result is a Bogoliubov-de Genne type of Hamiltonian
\be\lb{Hfermion}
H_{\rm fermion}= \sum_{k>0} [c_k^\dag,c_{-k}]\m{\e(k)}{\D(k)}{\D^*(k)}{-\e(-k)}\v{c_k}{c_{-k}^\dag},
\ee
where $\D(k)=i\cos(2\a)\sin k$ and $\e(k)=-\cos(k-2\a)$. In terms of Nambu spinor $\Psi^\dag_k\equiv [c_k^\dag,c_{-k}]$, $H_{\rm fermion}=\sum_{k>0}\Psi^\dag_k \hat{\mathcal{H}}(k)\Psi_k$, with $\hat{\mathcal{H}}(k)=d_0(k)\hat{I}+\sum_{i=x,y,z}d_i(k)\hat{\s}_i$, where $\hat{I}$ is the $2\times 2$ identity matrix and $\hat{\sigma}_{x,y,z}$ are the Pauli matrices. $d_0(k)=-\sin2\a\sin k$, $d_x(k)=0$, $d_y(k)=-\cos 2\a\sin k$ and $d_z(k)=-\cos 2\a\cos k$. The spectrum of fermion modes is given by $E_\pm(k)=-\sin(2\a)\sin k\pm |\cos(2\a)|$. The critical values for $\a$ where the spectrum $E_\pm(k)$ becomes gapless are given by $\a=\frac{1}{8}\pi,\frac{3}{8}\pi$. For $\a<\frac{1}{8}\pi$, the system is a $\hat{y}$-ferromagnet, while for $\a>\frac{3}{8}\pi$, it is a $\hat{y}$-anti-ferromagnet; Both phases are gapped. In the intermediate region, $\frac{1}{8}\pi<\a<\frac{3}{8}\pi$, it is in the $xy$-chiral phase with gapless excitations.

%-----------------------figure--------------------------------------------------
\begin{figure}[ht]
\begin{center}
\includegraphics[width=0.8 \textwidth]{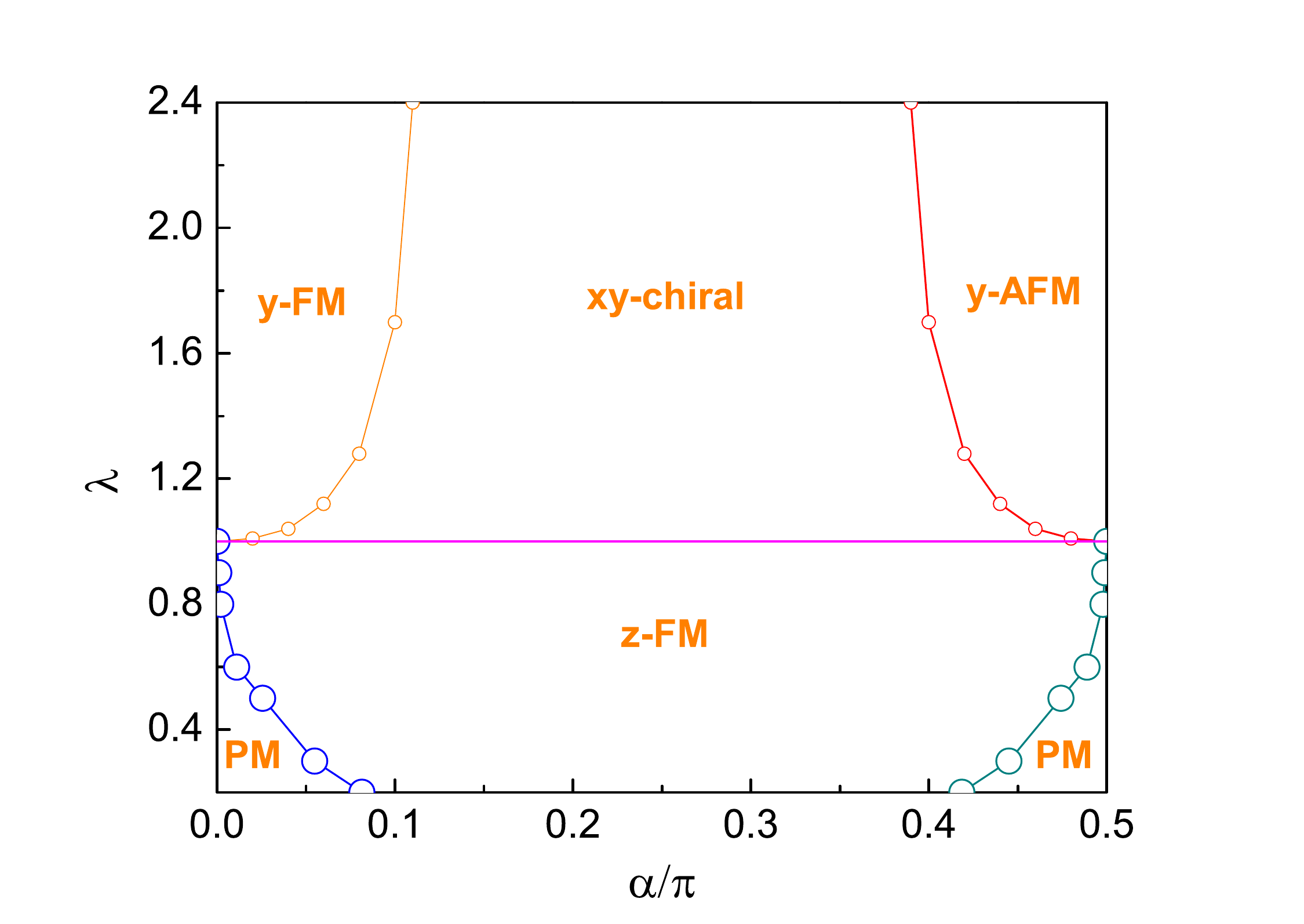}
\caption{Phase diagram of the effective spin model $H_{\rm mag}$ in the $\lambda$-$\alpha$ plane. In addition to the weak coupling magnetic phases, $\hat{z}$-FM and $xy$-chiral, one finds three additional magnetic phases: $\hat{y}$-FM, $\hat{y}$-AFM and PM states. Figure adapted from ref.\cite{Xu2014}}
\lb{1dmag}
\end{center}
\end{figure}
%-----------------------figure--------------------------------------------------

What is interesting is that the Hamiltonian eqn.(\ref{Hfermion}) describes $p$-wave pairing in one dimension, analogous to the Kitaev model~\cite{Kitaev2001}. The Hamiltonian obeys the following symmetry: $\hat{\mathcal{H}}(k)=-\s_x\hat{\mathcal{H}}(-k)^*\s_x$ and belongs to the ``D" symmetry class, characterised by a $\mathbb{Z}_2$ invariant~\cite{Ryu2010}. In the special case when $\a=0,\frac{\pi}{2}$, eqn.(\ref{Hfermion}) reduces to the standard Kitaev model. We note that the magnetic transitions described by JW fermions in the limit $\l\to\infty$ occur also for finite values of $\l>1$, as shown in Fig.\ref{1dsomag}. It is thus tempting to conclude that the phase boundaries between the $xy$-chiral and $\hat{y}$-ferromagnetic or $\hat{y}$-antiferromagnetic, to be described by the same topological transitions. Further investigations are necessary in this direction.
%-----------------------figure--------------------------------------------------
\begin{figure}[ht]
\begin{center}
\includegraphics[width=0.8 \textwidth]{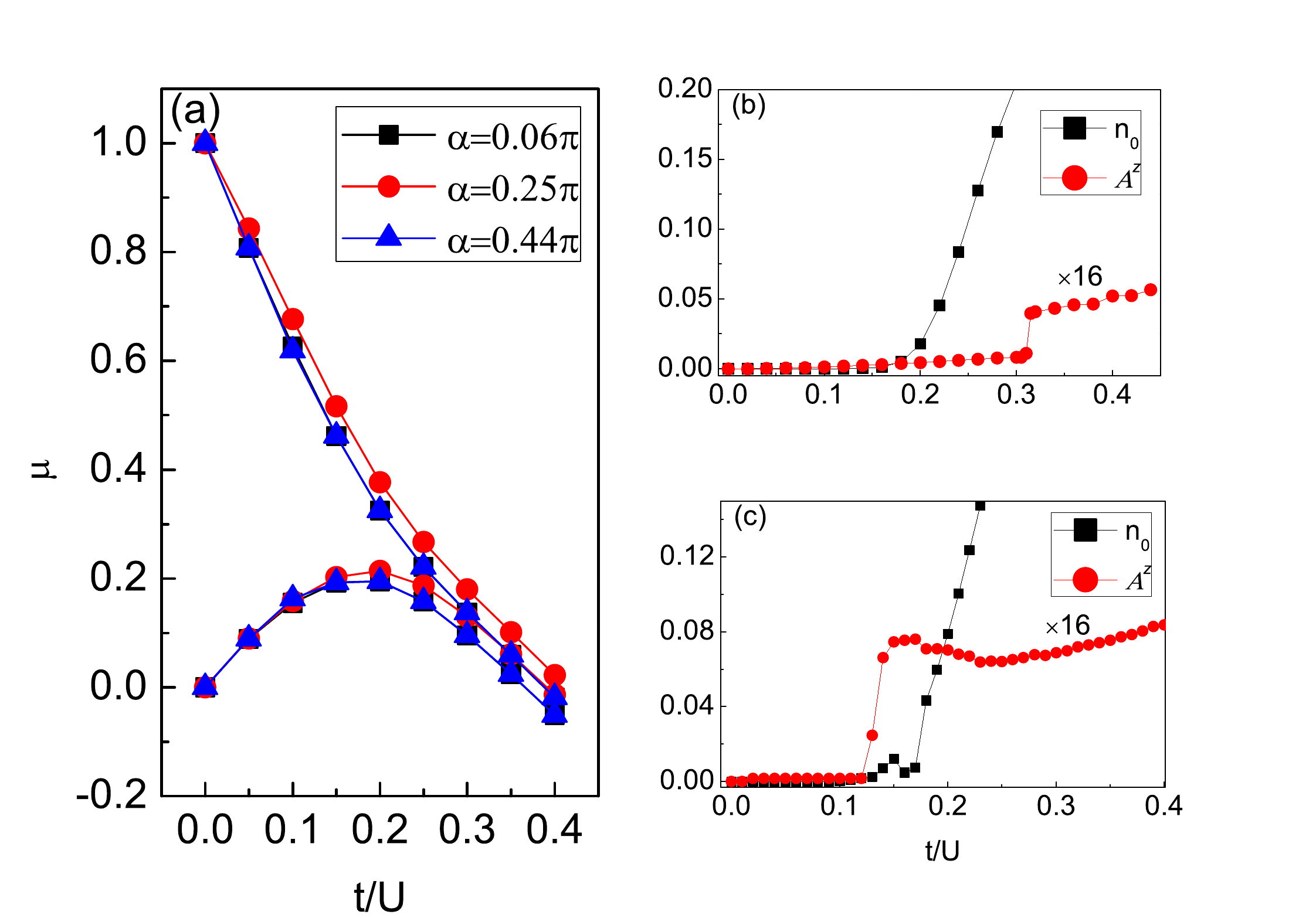}
\caption{(a) Phase diagram of the one-dimensional SOBHM in the $\mu$-$t/U$ plane. Note that for different values of $\a$, corresponding to various strength of spin-orbit coupling, the superfluid-Mott boundaries are only slightly modified. (b) and (c) show the magnetic transition from $\hat{y}$-FM to $xy$-chiral phase as one increase the hopping amplitude $t/U$. For (b) $\alpha=0.08\pi$ and $\lambda=1.5$ and for (c) $\alpha=0.07\pi$ and $\lambda=1.2$. $n_0$ can be regarded as the superfluid order parameter and describes the superfluid-Mott transition, while $A^z$ is the chiral order parameters. Note that the sequence of transition depends on the values of $\a$ and $\l$. Figure adapted from ref.\cite{Xu2014}}
\lb{1dsomag}
\end{center}
\end{figure}
%-----------------------figure--------------------------------------------------

Now, let us turn to the question of Mott-superfluid transition and in particular, how the magnetic phases obtained above evolve into the superfluid phase. For more detailed discussion, see refs~\cite{Zhao2014a,Zhao2014b,Xu2014}. It is instructive to look first at the weak coupling limit when $U\to 0$. In this case, the single particle spectrum has two degenerate ground states at $k=\pm \a$ with the corresponding wave function given by $\Psi_\pm(k)=\exp(\pm i a x)(1,\pm i)$. The superfluid order parameter is a superposition of the two states $\Psi_\pm(x)$ for $\l>1$, which leads to an order parameter of the form $(\lr{a_{x\ua}},\lr{a_{x\da}})=(\cos\a x,-\sin\a x)$. This corresponds to spin spiraling around the $\hat{y}$-axis with wave vector $2\a$. This is the $xy$-chiral phase found in the magnetic Hamiltonian, after rotating the spin around $\hat{x}$ by $\pi/2$. When $\l<1$, the system breaks the $\mathbb{Z}_2$ symmetry and chooses one of the $\Psi_\pm$ as its order parameter. The superfluid state is a ferromagnetic state along $\hat{y}$-direction, which, after rotating around $\hat{x}$-axis by $\pi/2$, is consistent with what is found in the strong coupling limit. 

However, strong interaction leads to more magnetic phases as is evident in the Mott regime, where additional paramagnetic, $\hat{y}$-ferromagnetic and $\hat{y}$-anti-ferromagnetic are found. The interesting question is whether these new magnetic phases, not found in the weak coupling limit, arise concomitant with emergence of Mott insulating phases, or they develop either before or after Mott-superfluid transition. To investigate this question, we first establish that the superfluid-Mott transition is only slightly modified by the presence of spin-orbit coupling. As an example, we calculate the $\mu$-$t/U$ phase diagram for three values of spin-orbit coupling $\a=0.06\pi, 0.25\pi, 0.44\pi$ by identifying values of $\mu$ and $t/U$ where single particle ($E_+\equiv E(N+1)-E(N)$) or hole excitation ($E_+\equiv E(N-1)-E(N)$) energies approach zero. As can be seen from Figure \ref{1dsomag}(a), the phase boundary is only slightly modified. 

On the other hand, the magnetic phases depend crucially on the value of spin-orbit coupling. In Figure \ref{1dsomag}(b,c), we show how magnetic phases in the strong coupling limit evolve into the superfluid phases, for two sets of parameters $(\a=0.08\pi,\l=1.5)$ and $(\a=0.07\pi,\l=1.2)$. We calculate the one-body density matrix $\langle a^\dag_{i\a}a_{j\s'}\rangle$ and extract its maximal eigenvalues $n_0$ whose eigenfunction decays algebraically. $n_0$ is non-zero only in the superfluid state. We also define the chiral correlation function $\mathcal{A}^\g(i,j)\equiv\langle A_i^{\gamma}A_j^{\gamma}\rangle$, where $\gamma=x,y,z$. In the Mott regime, $A_i^{\gamma}=\varepsilon^{\gamma \mu \nu}(s_i^{\mu}s_{i+1}^{\nu}-s_i^{\nu}s_{i+1}^{\mu})$, describing the chirality of the spins in the ground state, while in the superfluid state, we replace $s^\g_i=\frac{1}{2}a_{i\a}^\dag\s^\g_{\a\b}a_{i\b}$, with underlying boson operator. In the chiral state, one expects that the asymptotic value $\mathcal{A}^\g\equiv \lim_{|i-j|\to\infty}\mathcal{A}^\g(i,j)$ to remain finite. As can be seen from Figure \ref{1dsomag}(b,c), depending on the values of $(\a,\l)$, the magnetic transition can occur either before or after the superfluid transition. 

%In the case of fermions, ref Lin's work on flat band....

% resume WSC edits...
\subsubsection{Two-dimensional lattice with Rashba spin-orbit coupling}
Let us now turn to the two dimensional case with Rashba SOC. The Hamiltonian is given by Eqn.(\ref{sobhm}) with the hopping matrices given in Eqn.(\ref{hopx}) and Eqn.(\ref{hopy}). As before, we have $\a$, $t/U$ and $\l\equiv U'/U$ as three independent dimensionless parameters.

It was noted previously that the single-particle spectrum has four degenerate lowest energy states. As a result, any state where $N$ bosons are distributed among these minima is a valid ground state in the absence of interactions. It is expected that when weak interactions are taken into account, a unique ground state will be selected. To explore this, we first assume that the bosons condense into one single-particle state with the generic wavefunction $\varphi(\vec{r})=\sum_{m=1}^4 c_{{\bf k}_m} e^{i{\bf k}_m \cdot \vec{r}} \chi_{{\bf k}_m}$, where the $c_m$ are a set of normalized complex variational parameters and $\chi_{{\bf k}m}$ are the associated spin wave functions at the four minima. The optimal set of $c_m$ minimizes the interaction energy $E_{\rm int} [\{c_m\}] \equiv \bra{\Phi} H_{\rm int} \ket{\Phi}$, and fully characterizes the properties of the condensate. It is convenient to rewrite $H_{\rm int}$ in terms of the operators that diagonalize $H_0$, and then the interaction energy only receives contributions from terms where all 4 operators correspond to the 4 minima. This process yields an expression
\be
E_{\rm int} \propto \sum_{\alpha\beta} \sum_{kpq}{}^{'} U_{\alpha\beta} (c_{p+q-k} \chi_{p+q-k,\alpha}^{-})^*(c_{k} \chi_{k,\beta}^{-})^*(c_{p} \chi_{p,\beta}^{-})(c_{q} \chi_{q,\alpha}^{-}).
\ee
The primed sum indicating that we only consider terms where all momentum indices correspond to energy minima.

Minimizing this quantity, we find -- similar to studies in the absence of an optical lattice \cite{Wang2010,OZAWA2012a,OZAWA2012b,Hu2012} -- that either a single minimum is occupied (leading to a ``plane wave" condensate) or two opposite momenta are equally occupied (leading to a uniform density but spin-polarization-striped condensate). Which state is chosen depends on the deviation from a totally spin-isotropic interaction $U=U'$, with the striped condensate being energetically favorable when $U' > U$.

The conceptual explanation of this result is rather straightforward, but it is useful to first write down the wavefunctions which are macroscopically occupied. We have, for the plane wave phase,
\be
\Psi_{\rm PW}(\vec{r}) = \frac{1}{\sqrt{2}} \exp\left( i \vec{k}_1 \cdot \vec{r} \right) \left( \begin{array}{c} 1 \\ e^{i\pi/4} \end{array} \right).
\ee
For the stripe phase, 
\be
\Psi_{\rm stripe}(\vec{r}) = \frac{1}{2} \left[ \Psi_{{\rm PW},\vec{k}_1}(\vec{r}) + \Psi_{{\rm PW},\vec{k}_3}(\vec{r}) \right]
= \left( \begin{array}{c} \cos(\vec{k}_1 \cdot \vec{r}) \\  e^{i3\pi/4}\sin(\vec{k}_1 \cdot \vec{r}) \end{array} \right).
\ee
We may also consider, although it fails to appear as a ground state in this weak-coupling approach, a Skyrmion state where an equal-weight combination of all four minima is occupied,
\be
\Psi_{\rm Skyrmion}(\vec{r}) = \frac{1}{2\sqrt{2}} \sum_{m=1}^{4} \exp\left( i \vec{k}_m \cdot \vec{r} \right) \left( \begin{array}{c} 1 \\ e^{i(2m-1)\pi/4} \end{array} \right)
\ee
The stripe and plane wave solutions are the only two states that can be constructed in this way which have a spatially uniform number density, which is favored by the spin-isotropic part of the interaction. The stripe phase additionally has a sort of ``phase separation" into regions where the two spin densities minimize their spatial overlap. This is favored when the interspecies interaction $U'$ is dominant. The Skyrmion state describes a \emph{local} minimum in energy, but is never a global minimum. It does not have a uniform number density, and is less efficient than the stripe phase at minimizing the spatial overlap of the two spin components. It is interesting to note, however, that other authors have observed that density-modulated condensates, including quasicrystals, can be stabilized by \emph{long-ranged dipolar} interactions, even in the weak-coupling limit~\cite{wilson_PRL_2013, gopalakrishnan_PRL_2013}.

%-----------------------figure--------------------------------------------------
\begin{figure}[h]
\begin{center}
\includegraphics[width=1\textwidth]{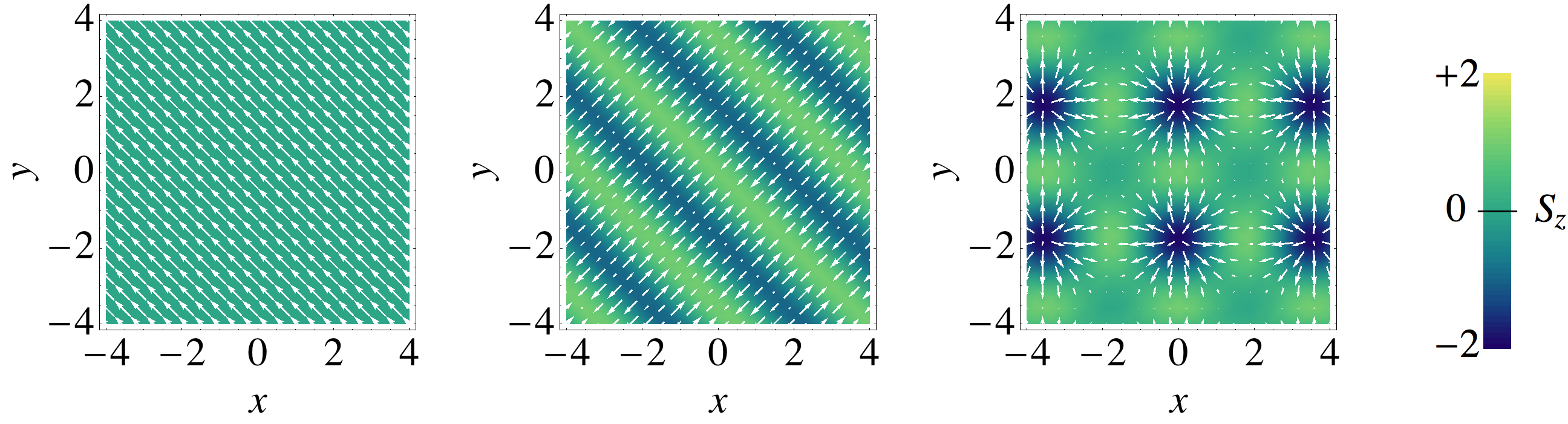}
\caption{Spin densities of the various ordered condensates: (left) single plane wave, (middle) stripe, (right) Skyrmion. The $z$ projection of the spin density is indicated by color, while the $x$ and $y$ projections are indicated by the white arrows. The plane wave and stripe solutions have uniform total number density, while the Skyrmion has a density wave, with peaks in the dark blue regions and vanishing density in the interstitial regions where all spin components are zero.}
\label{fig:gp_structure}
\end{center}
\end{figure}
%%-----------------------figure--------------------------------------------------

In the weakly interacting limit, the interactions are responsible for supporting a unique ground state and the structure of the ground state emerges from interference between the spinor wavefunctions describing the single-particle minima. In the opposite limit where $U,U' \gg t$, we begin, however, with the single-site spectrum of $H_{\rm int}$, as the band structure and low-energy states of $H_0$ are less relevant. In the following, we restrict our attention to unit filling. Then, reintroducing $H_0$ to second-order in perturbation theory yields the model given by Eq.~(\ref{eq:exchange}) and the paragraph that follows it.

These exchange interactions are frustrated even on the square lattice, and finding ground states is quite challenging. To gain some insight into the possible states supported by such a model, we revert to classical Monte Carlo simulations, in which we treat the spins $\vec{S}$ as classical variables. The resulting phase diagram is shown in the $\lambda\equiv U'/U$ and $\a$ plane in Fig.(\ref{fig:strongpd}), together with a few selected spin configurations. We characterize the different phases through the magnetic structure factor $S_{\vec{q}}=|\sum_x \vec{S}_x e^{i\vec{q}\cdot\vec{r}_x}|$. The peaks in this structure factor tell us about the magnetic ordering vectors (as shown in the inset of Fig.\ref{fig:strongpd}). A summary of various magnetic phases is given in Table \ref{tab:phases}, and we now proceed to point out some interesting features of the various phases.

(I) The existence of the two ferromagnetic phases which occupy the small $\a$ region can be understood as follows. In the limit $\a\to 0$, the magnetic Hamiltonian Eq.(\ref{eq:exchange}) reduces to the standard Heisenberg XXZ model, with the only anisotropy in the exchange coming from $\lambda \neq 1$. For $\lambda>1$, the $\hat{z}$-component of the exchange interaction is larger than the in-plane component, and one then expects ordering along $\hat{z}$. When $\lambda<1$, we have the opposite case. For small nonzero $\a$, these phases survive but with additional Ising anisotropies that pin the direction of the $xy$-ferromagnetism.

(II) The existence of the two magnetic phases near $\a=\pi/2$ can likewise be understood in the limiting case. The $\vec{D}$ vectors again vanish and the $\lambda$ dependence that leads to $z$-axis or $xy$-plane anisotropy is identical to the previous case. Now, though, the sign has switched so that the $z$-component of the exchange is antiferromagnetic. For the vortex crystal (VX) phase, the exchange along $x$ and $y$ directions are of different sign and additionally the exchange in spin space is of different sign for the $x$ and $y$ components. Ordering is therefore frustrated. To understand the classical phase that emerges here, a variational solution is useful. We propose a state $S_i^x=(-1)^{x_i} \sin\varphi$, $S_i^y = (-1)^{y_i}\cos\varphi$ with a uniform $\varphi$. Now, plugging this state into the full hamiltonian, we find that the energy is independent of $\varphi$; that is, the VX phase has a $U(1)$ degeneracy. For illustrative purposes, we settle on the choice $\varphi=\pi/4$, as this state emerges in our Monte Carlo annealing. We conjecture that this is because the degeneracy is broken slightly above $T=0$ by thermal and quantum fluctuations. Finally, because the state is coplanar, it gains no energy from the DM term. As $\a$ is reduced, the DM term grows and the coplanar state becomes unstable giving way to the non-coplanar Skyrmion crystal (SkX).

(III) For $\l >1$ and intermediate values of the spin-orbit coupling, we recover a magnetic phase reminiscent of the ``stripe" condensate described previously. Here we have an incommensurate spin spiral along the $(11)$ or $(1\bar{1})$ direction of the lattice. This sort of spiral state results quite generically from the combination of ferromagnetic exchange with any nonzero DM interaction. At weak coupling, the pitch of the stripe condensate was determined solely by the location of the energy minima $\vec{k}_m$, while at strong coupling, it is determined by the ratio of the DM interaction to the spin-isotropic interaction. Thus, even though the magnetic structure is similar, the underlying mechanism -- interference in the condensate and superexchange in the insulator -- is quite different. For $\l < 1$, coplanar spiral order is also found, but the spiral vector is along the $(10)$ or $(01)$ direction. This kind of order does the most to compromise between the DM term (tumbling the spins in one direction so that the cross product between spins in that direction does not vanish) while also satisfying the large compass interaction (by aligning spins along the other direction).

(IV) In a small parameter regime, we find that the energy is minimized by a superposition of stripes in the $(10)$ and $(01)$ directions. This superposition leads to a magnetic texture that again is reminiscent of the Skyrmion condensate described above. In this case it has a unit cell of $3 \times 3$ lattice sites. The central spin in the unit cell points in either the positive or negative $z$ direction, while the remaining spins tumble outward toward the opposite $z$ direction. For $\l < 1$, the extra planar anisotropy prevents these off-center spins from having a significant $z$-component, however. As the only non-coplanar arrangement of spins, this state also carries a non-zero spin chirality $\sum_{i} \vec{S}_i \cdot \left(\vec{S}_{i+\hat{x}} \times \vec{S}_{i+\hat{y}} \right)$. In this lattice discretization of the spin chirality, there is no need for the result to be quantized, however it is useful to note that this definition is inspired by a continuum formulation, wherein this spin chirality \emph{is} topologically quantized, and simply counts the number of Skyrmions present in the texture.

%-----------------------figure--------------------------------------------------
\begin{figure}[ht]
\begin{center}
\includegraphics[width=1 \textwidth]{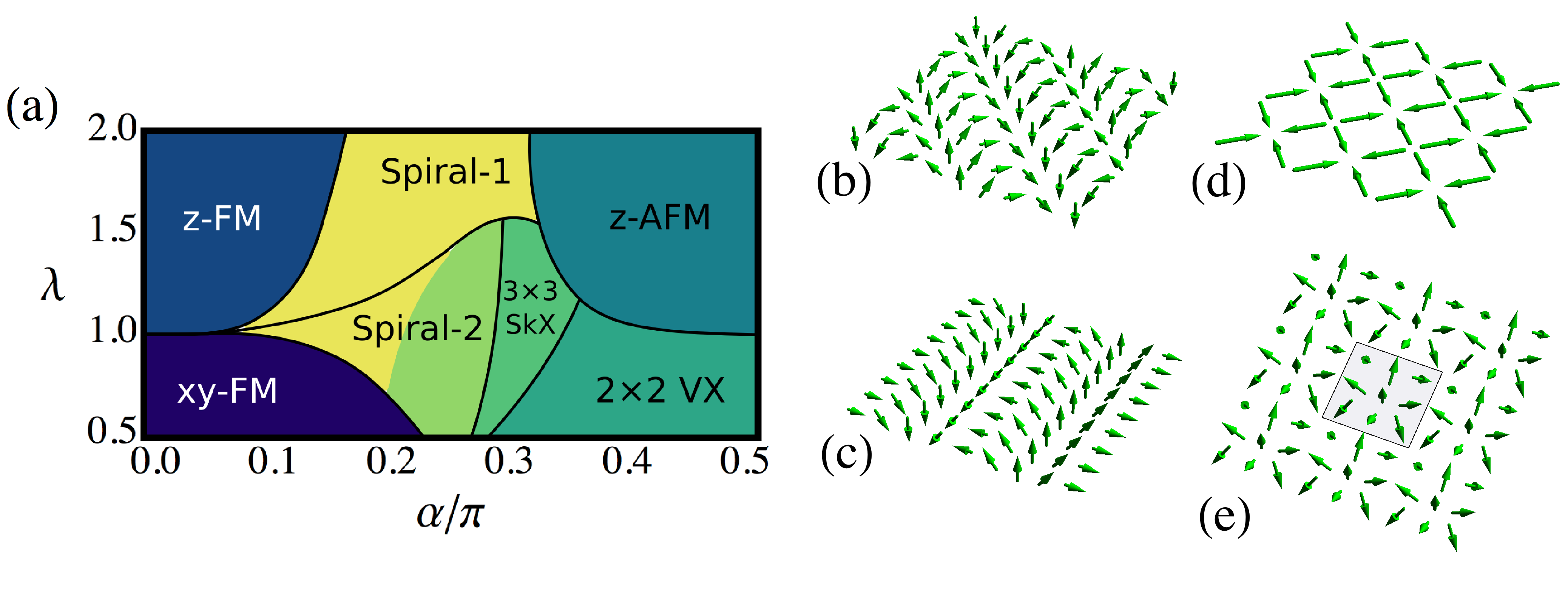}
\caption{(a) Magnetic phase diagram in the deep Mott limit determined by Monte Carlo annealing on a $36\times 36$ site square lattice. Although the spiral states are generally incommensurate, the shaded green area in the Spiral-2 region corresponds to a likely commensurate state. In this region the spiral unit cell contains 4 sites, with the spin winding by $\pi/2$ along each bond in the spiral direction. This pattern maximizes the cross-product of neighboring spins, and is therefore quite favorable in the region $\theta \sim \pi/4$, where the isotropic ferromagnetic interaction vanishes. On the right side, we show several classical spin configurations. (b) Spiral-1 state; coplanar spin-orientation rotating along (11). (c) Spiral-2 state; coplanar spin-orientation rotating along (10). (d) Vortex crystal; $2\times 2$ unit cell with $\pi/2$ rotation along each bond. (e) Skyrmion crystal; the $3\times 3$ unit cell is highlighted with a gray box. The central spin points in the positive $z$ direction, while the remaining spins tumble outward toward $-z$. Figure adapted from ref.\cite{Cole2012}}
\lb{fig:strongpd}
\end{center}
\end{figure}
%-----------------------figure--------------------------------------------------
%-----------------------table--------------------------------------------------
\begin{table}[h]
\begin{center}
\caption{\label{tab:phases}Summary of the classical spin states supported by the effective hamiltonian.}
\def\arraystretch{1.75}
{\setlength{\tabcolsep}{1em}
\begin{tabular}{| c | c | c |}
\hline
Phase & \pbox{6cm}{Location of peaks in \\$S_{\vec{q}}=|\sum_x \vec{S}_x e^{i\vec{q}\cdot\vec{r}_x}|$} & Spin orientation \\[1.5ex]
\hline
zFM & $(0,0)$ & along $z$ \\
xyFM & $(0,0)$ & \pbox{5cm}{in $xy$ plane, at angle \\ $(2 n + 1) \pi/4$ to the $x$ axis} \\
zAFM & $(\pi,\pi)$ & along $z$ \\
Spiral-1 & $(q,\pm q)$ & in $z$-$\vec{q}$ plane \\
Spiral-2 & $(q,0)$ or $(0,q)$ & in $z$-$\vec{q}$ plane \\
Vortex Crystal (VX) & $(\pi,0)$ and $(0,\pi)$ & \pbox{5cm}{in $xy$ plane, spin components: \\ $S_x=(-1)^x/\sqrt{2}$, $S_y = (-1)^y/\sqrt{2}$} \\
Skyrmion Crystal (SkX) & $(2\pi/3,0)$ and $(0,2\pi/3)$ & \pbox{5cm}{non-coplanar} \\
\hline
\end{tabular}}
\def\arraystretch{1}
\end{center}
\end{table}
%-----------------------table--------------------------------------------------

%
%
%\begin{table}
%{\begin{tabular}{@{}ccccccccccc@{}}
%\toprule\\[-6pt]
%		Phase & \pbox{6cm}{Location of peaks in \\$S_{\vec{q}}=|\sum_x \vec{S}_x e^{i\vec{q}\cdot\vec{r}_x}|$} & Spin orientation \\[1.5ex]
%		\hline
%		zFM & $(0,0)$ & along $z$ \\
%		xyFM & $(0,0)$ & \pbox{5cm}{\centering in $xy$ plane, at angle $(2 n + 1) \pi/4$ to the $x$ axis} \\
%		zAFM & $(\pi,\pi)$ & along $z$ \\
%		Spiral-1 & $(q,\pm q)$ & in $z$-$\vec{q}$ plane \\
%		Spiral-2 & $(q,0)$ or $(0,q)$ & in $z$-$\vec{q}$ plane \\
%		Vortex Crystal (VX) & $(\pi,0)$ and $(0,\pi)$ & \pbox{7cm}{\centering in $xy$ plane, spin components: $S^x=(-1)^{r_x}/\sqrt{2}$, $S^y = (-1)^{r_y}/\sqrt{2}$} \\
%		Skyrmion Crystal (SkX) & $(2\pi/3,0)$ and $(0,2\pi/3)$ & \pbox{5cm}{non-coplanar} \\
%		\hline\Hline
%\end{tabular}}
%\caption{Summary of the classical spin states supported by the effective hamiltonian.}
%\label{tab:phases}
%\end{table}

In between the two limits so far described, a transition must occur from a Mott insulating phase supporting various magnetic structures to a superfluid phase. Unlike the one-dimensional case, where exact numerical methods (DMRG, for example) can be applied, here we must resort to mean field theory. In the presence of spin-orbit coupling, the order-parameter is multi-component and may vary from site to site to incorporate inhomogeneous spin-density and phase structure. This requires us to extend the standard homogeneous Gutzwiller mean field theory for the Bose-Hubbard model to consider a more general Gutzwiller ansatz:
\begin{align}
\ket{\Psi} = \prod_i \bigg( & f_{i,0} + f_{i,1,1} b^\dagger_{i\su} + f_{i,1,-1} b^\dagger_{i\sd} + \nonumber \\
+ & f_{i,2,2} b^\dagger_{i\su} b^\dagger_{i\su} +  f_{i,2,0} b^\dagger_{i\sd} b^\dagger_{i\su} +  f_{i,2,-2} b^\dagger_{i\sd} b^\dagger_{i\sd} \ldots \bigg) \ket{0}
\label{eq:gutz_wf}
\end{align}
where the coefficients $\{ f_{i,2S,2m_S} \}$ can be determined by diagonalization in the local Hilbert space of each lattice site, where each site is coupled to its neighbors through the whole set of $\varphi_{i\sigma} \equiv \lr{b_{i\s}}$, and the set of coefficients are sought which satisfy the self-consistency condition at each site. By starting with different initial
states we can then search for global energy minima in the space of such self-consistent solutions. From Eq.~\ref{eq:gutz_wf}, it is clear that any number of terms can be added (the local Hilbert space is infinite), but to study the $n=1$ Mott insulator to superfluid transition, the six terms written are typically sufficient since higher occupancies are strongly
suppressed by the Hubbard repulsion near the Mott transition. Finally, due to the inhomogeneity expected in the solution, calculations must be carried out on a finite cluster of linear 
dimension $L$, checking for the stability of the ground state as $L$ is varied.

Similar questions about the Mott-superfluid transition in two dimensions arise as those from the one-dimensional case. (1) Does the superexchange-induced magnetic order in the insulator persist across the Mott transition? (2) If so, how does this impact the nature of the transition, compared to the traditional Bose-Hubbard model? (3) Just above the Mott transition, can there exist spin-ordered superfluid phases which have no weak-coupling analog?

\begin{figure}[t]
 \centering
 \includegraphics[width=0.8\textwidth]{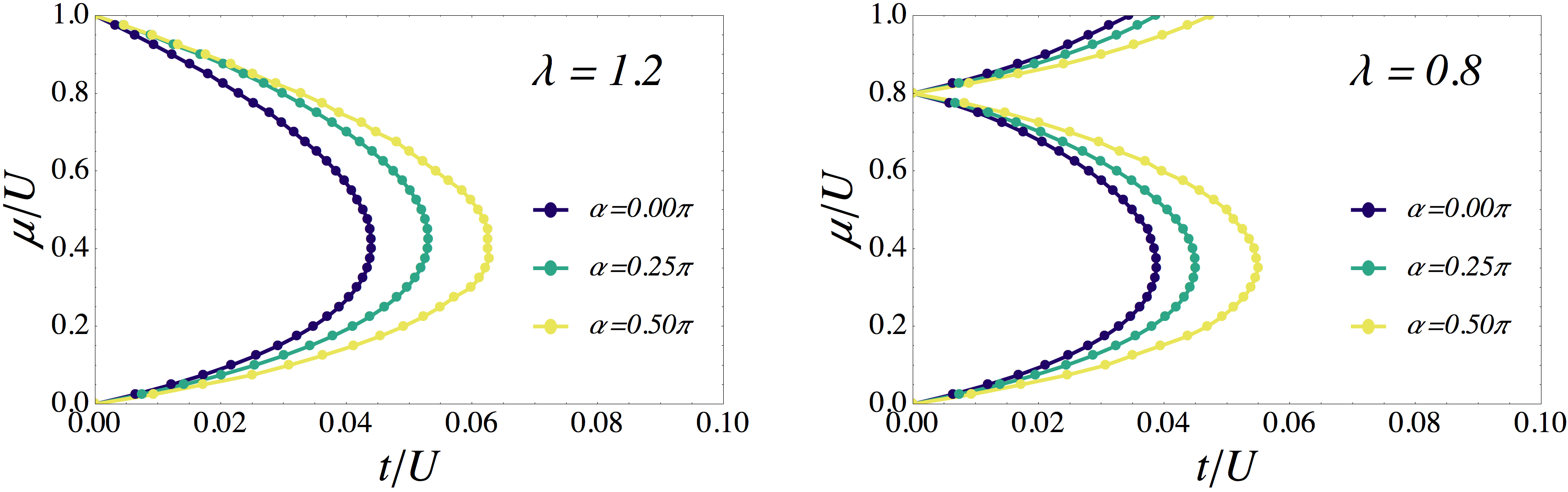}
\caption{
(Color online) Phase diagrams of the spin-orbit coupled Bose-Hubbard model in $\mu/U$ {\it vs}. $t/U$ plane, showing Mott lobes and superfluid states. (A) phase diagram with $\lambda=1.2$ and $\alpha=(0, 0.25, 0.5)\pi$ and (B) $\lambda=0.8$ and $\alpha=(0, 0.25, 0.5)\pi$. The width of the $n=1$ lobe is given by $\lambda U$ and the critical value $(t/U)_c$ increases with $\lambda$.}
\label{fig:phase-diagram}
\end{figure}

Some representative results of the Gutzwiller approach are shown in Fig.~\ref{fig:phase-diagram}, where we plot the Mott lobes for filling $n=1$ and $\alpha=0,\pi/4,\pi/2$. In general, increasing $\alpha$ frustrates the hopping so that a larger bare $t$ is required to support a superfluid phase. This result was first obtained, absent the possibility of a spatially varying order parameter, by Gra\ss, et al. using a series expansion approach~\cite{grass_PRA_2011}. This series expansion method has also been used to give some indication of the excitations of the model~\cite{mandal_PRB_2012}, as well as the very intriguing proposal by Wong and Duine~\cite{Wong2013} of the possibility of quasiparticle excitation bands that carry nontrivial Chern number, which will be discussed shortly.

\begin{figure}[t]
 \centering
 \includegraphics[width=0.8\textwidth]{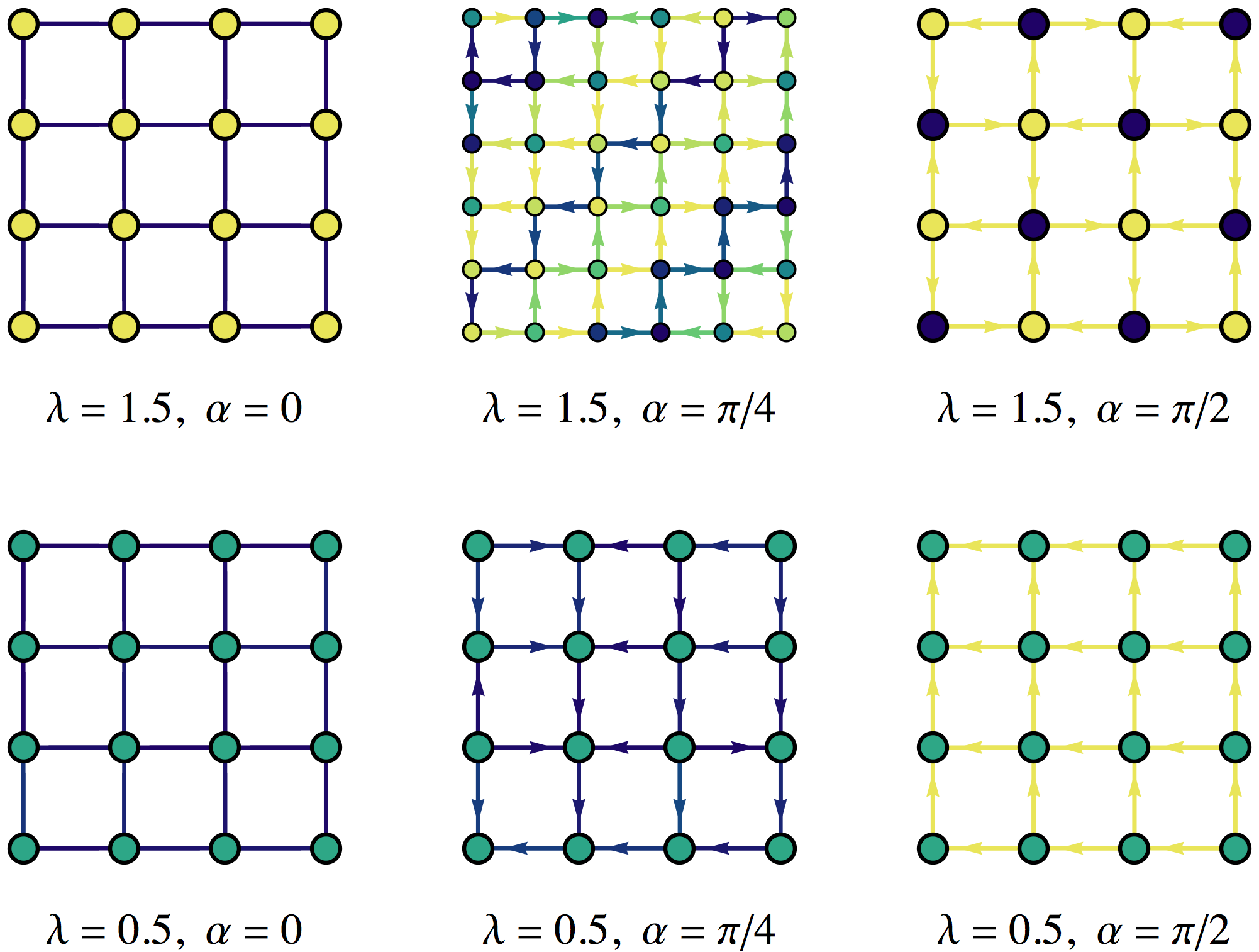}
\caption{
(Color online) Magnetic structure and currents in the strong coupled superfluid state close to the Mott-superfluid phase boundary. Blue (yellow) dots denote spin up (down), while green ones indicate that the magnetization is ordered in the $xy$-plane. Number currents are plotted as arrows on the bonds, with the brightness of the color indicating the magnitude of the current.}
\label{fig:2DSFStructure}
\end{figure}

In addition to merely locating the phase transition, the Gutzwiller approach also allows for a spatially varying order-parameter, and it is interesting to investigate the strong-coupling superfluid states by looking at the spatial dependence of the solutions. One way to characterize these states is through local spin-densities and bond currents
\be
{\bf m}_{i}=\lr{b^\dag_{i\mu}\boldsymbol{\s}_{\mu\mu'}b_{i\mu'}}.
\ee
\be
J_{ij}^{\mu\nu}= -it(\mathcal{R}_{ij}^{\mu\nu} \langle b_{i\mu}^\dag b_{j\nu} \rangle - c.c.)
\ee
The latter quantity describes the current flow from site $j$ to $i$, with $\mu\nu$ indicating that it is a tensor in the spin space. In Figure \ref{fig:2DSFStructure}, we show the $\hat{z}$-component of the magnetization ${\bf m}_{iz}$ and the number current along the bond $\tilde{J}_{ij}=\sum_\mu J_{ij}^{\mu\mu}$ for a variety of mean-field states. All of these states have uniform number density, but exhibit different magnetic order. For example, with $\l=1.5$ and $\a=\pi/2$, the superfluid state exhibits zAF magnetic order, consistent with the magnetic phase in the Mott insulating regime, while at $\l=0.5$ the magnetization is in the plane and adopts the VX structure. In addition, however, plaquette currents are generated in the superfluid, which can be understood from a slave boson construction of the SOBHM, which we shall not discuss here~\cite{Cole2012}. We note that such current patterns can be observed in experiments using quantum quenches~\cite{Killi2012}. Similar results to those outlined here have also been obtained by other authors at this level of approximation~\cite{qian_arXiv_2013}. Additionally, recent calculations using a much more sophisticated bosonic DMFT approach have led to similar conclusions~\cite{he_arXiv_2014}.

Insights into the nature of the bond currents in the ground state may be obtained using a slave boson approach, which has been formulated for spinor bosons
in the context of the SOBHM\cite{Cole2012}. For magnetically ordered superfluids, this approach is particularly simple to understand, and it amounts to freezing the 
spinor part of the boson wavefunction
while allowing for superfluidity and currents to be determined by the charge sector of the Hamiltonian. Schematically, we can set $b^\dag_{i\mu} = a^\dag_i z_{i\mu}$ where
the spinor wavefunction $z_{i\mu}$ is chosen to correspond to the spin structure of the ground state, and the $a$-boson simply carries a `charge' quantum number. Explicitly,
denoting angles $(\theta_i,\phi_i)$ to refer to the local spin direction, we arrive at
\begin{eqnarray}
z_{i\uparrow} &=& \cos(\theta_i/2) {\rm e}^{-i\phi_i/2} \\
z_{i\downarrow} &=& \sin(\theta_i/2) {\rm e}^{+i\phi_i/2},
\end{eqnarray}
so that the effective Hamiltonian for the $a$-bosons takes the form,
\begin{eqnarray}
H_a &=& - t \! \sum_{i\delta} (a^\dag_i a_{i+\delta} [z^*_{i\alpha} R_{i,i+\delta}^{\alpha\beta} 
z_{i+\delta,\beta}]
+ {\rm h.c.}) +  \frac{U}{2} \! \sum_i a^\dag_i a^\dag_i a_i a_i \notag\\
&+& (\lambda-1) U \sum_i |z_{i\uparrow}|^2 
 |z_{i\downarrow}|^2 a^\dag_i a^\dag_i a_i a_i.
 \label{Eq:Hbz}
\end{eqnarray}
It then becomes clear that $[z^*_{i\alpha} R_{i,i+\delta}^{\alpha\beta} z_{i+\delta,\beta}] \sim {\rm e}^{i {\cal A}_{i,i+\delta}} $ leads to an effective $U(1)$ gauge field 
(a `synthetic magnetic field') for the charge bosons, and different magnetic orders imprint different background gauge fields, which allows us to understand the 
novel bond current patterns found in the mean field theory. For instance, an Ising antiferromagnetic order, observed at $\alpha=\pi/2$ and $\lambda > 1$, 
imprints a $\pi$-flux through each plaquette for the
$a$-bosons, leading to a checkerboard pattern of current order, spontaneously breaking the translationally symmetry of the lattice as shown in Fig.\ref{fig:2DSFStructure}. 
Furthermore, the Hamiltonian in Eq.~(\ref{Eq:Hbz}) also allows us to describe the superfluid to Mott transition of bosons with a given magnetic order, as varying $U$ can lead to 
Mott localization of the $a$-bosons which describes charge localization in the SOBHM.

%-----------------------figure--------------------------------------------------
%-----------------------figure--------------------------------------------------

%-----------------------figure--------------------------------------------------
%-----------------------figure--------------------------------------------------

Finally, let us discuss the possible topological phases in the spin-orbit coupled Bose-Hubbard models, as proposed by Wong and Duine~\cite{Wong2013}. Unlike the Fermi system, where the topology of the band structure can be readily explored with free fermions, free bosons, even in a nontrivial band, will automatically condense into the lowest single particle state. Free bosons are thus insensitive to the topology of the band structure. In addition, for weak interactions, the Bose-condensed system exhibits gapless bulk phonon excitations, in contrast to the bulk-gapped topological insulators. As a result, a natural place to look for the possible emergence of topological properties for SOBHM is in the Mott insulating regime, where the single particle excitations are gapped due to strong interactions. 

For simplicity, let us consider a ferromagnetic Mott insulating state, as done in ref~\cite{Wong2013}. This simplifies the discussion considerably, as the Mott insulating state respects the lattice translational invariance. As a result, we can write the inverse of the single particle Green function in the Mott regime as
\be
-\hat{G}^{-1}(\o,{\bf k})=\tilde{d}_0(\o,{\bf k})+\tilde{d}_i(\omega,{\bf k})\cdot\boldsymbol{\sigma}.
\ee
The quasi-particle (hole) excitations are determined by $\det[\hat{G}^{-1}(\o,{\bf k})]=0$ and we shall denote the excitation energy as $\o_0({\bf k})$. The quasi-particle can then be regarded as moving in an effective magnetic field with an effective ${\bf d}$-vector given by
\be
\tilde{d}_i(\o_0({\bf k}),{\bf k}).
\ee
In the ground state, all the quasi-hole excitations are occupied and the topological character of the Mott insulating state can be determined by integrating the Berry curvature of the quasi-hole excitations over the Brillouin Zone. Within the random phase approximation for the SOBHM, the quasi-particle ${\bf \tilde{d}}$ assumes a particularly simple form
\begin{align}\nn
\tilde{d}_x(\o_0({\bf k}),{\bf k}) &=d_x({\bf k})\\\nn
\tilde{d}_y(\o_0({\bf k}),{\bf k}) &=d_y({\bf k})\\\nn
\tilde{d}_z(\o_0({\bf k}),{\bf k}) &=d_z({\bf k})+\frac{1}{2}(g^{-1}_{\da\da}(\o_0({\bf k}))-g^{-1}_{\ua\ua}(\o_0({\bf k})))
\end{align}
where $g_{\s\s}(\o)$ is the onsite Green function. Thus interactions enter only through the modification of the $\hat{z}$-component of the effective magnetic field $\boldsymbol{d}$. In the cases investigated in ref~\cite{Wong2013}, such a modification can lead to an integer Hall conductivity even though the underlying free-particle band structure is trivial. In the
language of the slave-boson picture discussed above, this means the spontaneous ordering of spins in the Mott insulator leads to time-reversal breaking, and an extra
added charge boson (particle or hole) in this case, which senses local orbital magnetic fields on plaquettes, can develop a gapped, topologically nontrivial, band structures 
with nonzero Chern numbers.

\section{Future prospects}
\lb{sec:prospect}
In this article, we first reviewed several experimental schemes currently employed to study the spin-orbit coupling in cold atoms in the continuum as well as in optical lattices. In the later case, we concentrate on the interplay between the strong onsite interaction and the spin-orbit coupling and point out several novel phenomena associated with them. Theoretically, a few outstanding questions remain to be understood in SOBHM.
\begin{itemize}
\item What is the nature of the superfluid to Mott insulator phase transitions in the presence of spin-orbit coupling? How does the density or spin modulation in the superfluid state
modify the critical properties of the transition?
\item It is necessary to better characterize the superfluid state by, for example, calculating and measuring the superfluid and spin superfluid density. The same problem remains 
to be done in the case of spin-orbit coupled quantum gases in the absence of an optical lattice.
\item Investigate the possible band topology and edge states in the Mott insulating regime where interesting magnetic phases (SkX and VX) are present.  
\end{itemize}
So far, only spin-orbit coupling along one direction is realized in actual experiments and there exists proposals to engineer Rashba spin-orbit coupling using extension of Raman scheme~\cite{Campbell2011,Sau2011,Xu2012} and pulsed inhomogeneous magnetic fields using atomic chip~\cite{Anderson2013}. The realisation of Rashba spin-orbit coupling would enable the study of the remarkable spin-textured Mott insulators 
and superfluids as unveiled theoretically.

Currently, the major obstacle with the Raman scheme is heating due to spontaneous emission and this concern seems to be less severe in the case of shaking lattice. The problem of heating could be mitigated by using atoms with a long-lived electronic excited state such as Yb~\cite{Fukuhara2007,Taie2012,Scazza2014,Cappellini2014} and Sr~\cite{Zhang2014}, or Lanthanide atoms like Dy~\cite{Lu2012} and Er~\cite{Aikawa2012}. These atoms offer, in addition to the possible spin-orbit coupling induced by Raman lasers, a larger manifold of spin states which could open the gateway towards new exotic quantum states in cold atoms~\cite{Cui2013}. On the other hand, with the recently realized Haldane model using shaking lattice~\cite{Jotzu2014}, an obvious next step would be to investigate the interaction effects and search for fractional Chern insulators~\cite{Neupert2011}. 

Finally, there may be interesting directions to explore by
putting spinor atoms in close proximity to the surfaces of cryogenic materials \cite{Lev2013}. Letting the atoms interact with surfaces of complex oxides
which can support novel magnetic textures may lead to novel gauge field configurations.

\section{Acknowlegement}
We would like to thank Xu Zhihao and Subroto Mukerjee for discussions. S.Z. is supported by a startup grant from the University of Hong Kong and a grant from the Research Grants Council of the Hong Kong Special Administrative Region, China (Grant No. HKUST3/CRF/13G).  WSC acknowledges NSF grant DMR139461 and NT was supported under ARO Grant No. W911NF-13-1-0018 with funds from the DARPA OLE program. AP 
acknowledges support from NSERC of Canada, and thanks the Aspen Center for Physics (Grant No. NSF PHY-1066293) for hospitaliy during completion of this manuscript.

\appendix

%******************************************
\section{Effective magnetic hamiltonian derived from two-site perturbation theory}
\label{app:strong_coupling}
%******************************************
Taking the limit $U,U' \gg t$ for the hamiltonian Eq.~(\ref{sobhm}), particle number fluctuations are effectively blocked, leaving only \emph{virtual} hopping processes to reduce the ground-state degeneracy. An effective magnetic hamiltonian governing these residual spin fluctuations can be derived for the Mott insulator by ordinary second-order perturbation theory. The starting point is to consider a restriction to two lattice sites, and to decompose the hamiltonian as $H=H_0+H_1$ where $H_0$ includes only the onsite (i.e., the interaction) terms and
\be
H_1= \sum_{\alpha\beta} \bdag_{1\alpha} h_{\alpha\beta} \bbbb_{2\beta} + \mbox{h.c.}
\ee

On these two sites, the set of lowest energy eigenstates of $H_0$ is spanned by
\be
\mathcal{G} = \left\{ \ket{\su_1,\su_2}, \ket{\su_1,\sd_2}, \ket{\sd_1,\su_2}, \ket{\sd_1,\sd_2} \right\}
\ee
which all have eigenvalue $-2\mu$. The calculation of perturbative shifts to the eigenvalues and eigenstates can be combined into an \emph{effective} magnetic hamiltonian in this basis, whose matrix elements between states $\ket{s_1},\ket{s_2} \in \mathcal{G}$ are
\be
(H_{\rm mag})_{s_1 s_2} = - \sum_{\gamma} \frac{\bra{s_1} H_1 \ket{\gamma} \bra{\gamma} H_1 \ket{s_2}}{E_{\gamma} - \frac{1}{2}\left( E_{s_1}+E_{s_2} \right)}
\label{eq:perturbation_hamiltonian}
\ee
where $\ket{\gamma}$ are the six available excited states which can be obtained by acting on the four ground states with $H_1$. The algebra that goes into constructing this spin model is tedious, but is included here for completeness. To start, we can calculate
\begin{align}
H_1 \ket{\mu_1,\nu_2} &=  \sum_{\alpha\beta} \left(
\bdag_{1\alpha} h_{\alpha\beta} \bbbb_{2\beta} + \bdag_{2\beta} h^*_{\alpha\beta} \bbbb_{1\alpha}
\right) \bdag_{1\mu} \bdag_{2\nu} \ket{0} \\
&=  \sum_{\alpha\beta} \left( 
\bdag_{1\alpha} h_{\alpha\beta} \bdag_{1\mu} \delta_{\beta,\nu} + \bdag_{2\beta} h^*_{\alpha\beta} \bdag_{2\nu} \delta_{\alpha,\mu}
\right)\ket{0} \\
&= \sum_{\alpha} h_{\alpha\nu} \bdag_{1\alpha} \bdag_{1\mu} \ket{0} +
 \sum_{\beta} h^*_{\mu\beta} \bdag_{2\beta} \bdag_{2\nu} \ket{0} 
\end{align}

While substituting in the four ground-state spin orientations yields
\begin{align}
H_1 \ket{\su_1,\su_2} =
&\sqrt{2} h_{\su\su} \ket{(\su\su)_1,0_2} + h_{\sd\su} \ket{(\su\sd)_1,0_2} + \nonumber \\
&\sqrt{2} h^*_{\su\su} \ket{0_1,(\su\su)_2} + h^*_{\su\sd} \ket{0_1,(\su\sd)_2} \\
H_1 \ket{\su_1,\sd_2} =
&\sqrt{2} h_{\su\sd} \ket{(\su\su)_1,0_2} + h_{\sd\sd} \ket{(\su\sd)_1,0_2} + \nonumber \\
&\sqrt{2} h^*_{\su\sd} \ket{0_1,(\sd\sd)_2} + h^*_{\su\su} \ket{0_1,(\su\sd)_2} \\
H_1 \ket{\sd_1,\su_2} =
&\sqrt{2} h_{\sd\su} \ket{(\sd\sd)_1,0_2} + h_{\su\su} \ket{(\su\sd)_1,0_2} + \nonumber \\
&\sqrt{2} h^*_{\sd\su} \ket{0_1,(\su\su)_2} + h^*_{\sd\sd} \ket{0_1,(\su\sd)_2} \\
H_1 \ket{\sd_1,\sd_2} =
&\sqrt{2} h_{\sd\sd} \ket{(\sd\sd)_1,0_2} + h_{\su\sd} \ket{(\su\sd)_1,0_2} + \nonumber \\
&\sqrt{2} h^*_{\sd\sd} \ket{0_1,(\sd\sd)_2} + h^*_{\sd\su} \ket{0_1,(\su\sd)_2}
\end{align}
The factors of $\sqrt{2}$ arise from normalization, e.g., $\ket{(\su\su)_1,0_2} = \frac{1}{\sqrt{2}} \bdag_{1\su} \bdag_{1\su} \ket{0}$. At this point it is convenient to rewrite this as a table of the matrix elements which go into Eq.~(\ref{eq:perturbation_hamiltonian}), provided in Table~\ref{tab:matrix_elements}.

\begin{table}[h]
\begin{center}
\begin{tabular}{|c|cccc||c|}
  \hline
  $\bra{\gamma} H_1 \ket{s}$ & $\ket{\su_1,\su_2}$ & $\ket{\su_1,\sd_2}$ & $\ket{\sd_1,\su_2}$ & $\ket{\sd_1,\sd_2}$ & $E_{\gamma} - \frac{1}{2}\left( E_{s_1}+E_{s_2} \right)$ \\
  \hline
  $\bra{(\su\su)_1,0_2}$ & $\sqrt{2}h_{\su\su}$		& $\sqrt{2}h_{\su\sd}$	& $0$				& $0$				& $U_{\su\su}$ \\
  $\bra{(\su\sd)_1,0_2}$ & $h_{\sd\su}$			& $h_{\sd\sd}$			& $h_{\su\su}$			& $h_{\su\sd}$			& $U_{\su\sd}$ \\
  $\bra{(\sd\sd)_1,0_2}$ & $0$					& $0$				& $\sqrt{2}h_{\sd\su}$	& $\sqrt{2}h_{\sd\sd}$	& $U_{\sd\sd}$ \\
  $\bra{0_1,(\su\su)_2}$ & $\sqrt{2}h^*_{\su\su}$	& $0$				& $\sqrt{2}h^*_{\sd\su}$	& $0$				& $U_{\su\su}$ \\
  $\bra{0_1,(\su\sd)_2}$ & $h^*_{\su\sd}$			& $h^*_{\su\su}$		& $h^*_{\sd\sd}$		& $h^*_{\sd\su}$		& $U_{\su\sd}$ \\
  $\bra{0_1,(\sd\sd)_2}$ & $0$					& $\sqrt{2}h^*_{\su\sd}$	& $0$				& $\sqrt{2}h^*_{\sd\sd}$	& $U_{\sd\sd}$ \\
  \hline
\end{tabular}
\caption{Virtual state matrix elements which enter into Eq.~(\ref{eq:perturbation_hamiltonian}) for calculating the low-energy effective spin hamiltonian deep in the Mott insulating limit. The rightmost column gives the energy gap to the corresponding virtual excitation.}
\label{tab:matrix_elements}
\end{center}
\end{table}
  
At this point we may insert the matrix elements in Table~\ref{tab:matrix_elements} into the expression Eq.~\ref{eq:perturbation_hamiltonian}. The result is a $4 \times 4$ matrix of couplings which is not particularly illuminating. However we will need these matrix elements for an alternative representation with a more physical character, which we now construct.

Each term in the effective hamiltonian can be replaced by a boson operator expression, since $H_{\rm mag}$ can be expanded in the set of projection operators into $\mathcal{G}$
\be
\ket{\sigma_1\sigma'_2} \bra{\tau_1\tau'_2} = \bdag_{1\sigma} \bdag_{2\sigma'} \bbbb_{2\tau'} \bbbb_{1\tau}
\ee
which can then be expressed in local spin operators through the transformations
\be
\bdag_{i\su} \bbbb_{i\su} = \frac{1}{2} + S_i^z, \quad
\bdag_{i\sd} \bbbb_{i\sd} = \frac{1}{2} - S_i^z, \quad
\bdag_{i\su} \bbbb_{i\sd} = S_i^+, \quad
\bdag_{i\sd} \bbbb_{i\su} = S_i^-
\ee

In the remainder of this Appendix, we carry out the expansion of $H_{\rm mag}$ and rearrangement of terms that give the more familiar magnetic model presented in the main text. First the expansion, which is simply writing the sum $H_{\rm mag} = \sum_{\alpha\beta} (H_{\rm mag})_{\alpha\beta}\ket{\alpha}\bra{\beta}$ out explicitly. We simplify the notation by taking $V_{\alpha\beta} \equiv \left(H_{\rm mag}\right)_{\alpha\beta}$
\begin{align}
H_{\rm mag} = \;
& V_{11} \bdag_{1\su}  \bbbb_{1\su} \bdag_{2\su} \bbbb_{2\su} +
V_{22} \bdag_{1\su}  \bbbb_{1\su} \bdag_{2\sd} \bbbb_{2\sd} + \nonumber \\
& V_{33} \bdag_{1\sd}  \bbbb_{1\sd} \bdag_{2\su} \bbbb_{2\su} +
V_{44} \bdag_{1\sd}  \bbbb_{1\sd} \bdag_{2\sd} \bbbb_{2\sd} + \nonumber \\ \bigg(
& V_{12} \bdag_{1\su}  \bbbb_{1\su} \bdag_{2\su} \bbbb_{2\sd} +
V_{13} \bdag_{1\su}  \bbbb_{1\sd} \bdag_{2\su} \bbbb_{2\su} + \nonumber \\
& V_{14} \bdag_{1\su}  \bbbb_{1\sd} \bdag_{2\su} \bbbb_{2\sd} +
V_{23} \bdag_{1\su}  \bbbb_{1\sd} \bdag_{2\sd} \bbbb_{2\su} + \nonumber \\
& V_{24} \bdag_{1\su}  \bbbb_{1\sd} \bdag_{2\sd} \bbbb_{2\sd} +
V_{34} \bdag_{1\sd}  \bbbb_{1\sd} \bdag_{2\su} \bbbb_{2\sd}
+ \mbox{h.c.} \bigg)
\end{align}

Now we insert the substitution of spin operators
\begin{align}
H_{\rm mag} = \;
& V_{11} \left( \frac{1}{2} + S_1^z \right) \left( \frac{1}{2} + S_2^z \right) +
V_{22} \left( \frac{1}{2} + S_1^z \right) \left( \frac{1}{2} - S_2^z \right) + \nonumber \\
& V_{33} \left( \frac{1}{2} - S_1^z \right) \left( \frac{1}{2} + S_2^z \right) +
V_{44} \left( \frac{1}{2} - S_1^z \right) \left( \frac{1}{2} - S_2^z \right) + \nonumber \\ \bigg(
& V_{12} \left( \frac{1}{2} + S_1^z \right) S_2^+ +
V_{13} S_1^+ \left( \frac{1}{2} + S_2^z \right) + V_{14} S_1^+ S_2^+ + \nonumber \\
& V_{23} S_1^+ S_2^- + V_{24} S_1^+ \left( \frac{1}{2} - S_2^z \right) +
V_{34} \left( \frac{1}{2} - S_1^z \right) S_2^+
+ \mbox{h.c.} \bigg)
\end{align}

Next we begin the process of arranging these terms
\begin{align}
H_{\rm mag} = \;
& \frac{1}{4} \left( V_{11} + V_{22} + V_{33} + V_{44} \right) + \left( V_{11} - V_{22} - V_{33} + V_{44} \right) S_1^z S_2^z + \nonumber \\
& \frac{1}{2} \left( V_{11} + V_{22} - V_{33} - V_{44} \right) S_1^z + \frac{1}{2} \left( V_{11} - V_{22} + V_{33} - V_{44} \right) S_2^z + \nonumber \\
& \frac{1}{2} \big[ (V_{13}+V_{24}) S_1^+ + (V_{12}+V_{34}) S_2^+ +  \nonumber \\ & (V_{13}+V_{24})^* S_1^- + (V_{12}+V_{34})^* S_2^- \big] + \nonumber \\
& (V_{12}-V_{34}) S_1^z S_2^+ + (V_{13}-V_{24}) S_1^+ S_2^z + \nonumber \\ & (V_{12}-V_{34})^* S_1^z S_2^- + (V_{13}-V_{24})^* S_1^- S_2^z + \nonumber \\
& V_{14} S_1^+ S_2^+ + V_{23} S_1^+ S_2^- + V_{14}^* S_1^- S_2^- + V_{23}^* S_1^- S_2^+
\end{align}

The spin raising and lowering operators are convenient for many purposes, but here we revert back to the cartesian components through $S^{\pm}=S^x \pm iS^y$, and, throwing out the overall constant term, we write
\begin{align}
H_{\rm mag} = \;
& \left( V_{11} - V_{22} - V_{33} + V_{44} \right) S_1^z S_2^z + \nonumber \\
& \frac{1}{2} \left( V_{11} + V_{22} - V_{33} - V_{44} \right) S_1^z + \frac{1}{2} \left( V_{11} - V_{22} + V_{33} - V_{44} \right) S_2^z + \nonumber \\
& \Re\left(V_{13}+V_{24}\right) S_1^x + \Re\left(V_{12}+V_{34}\right) S_2^x - \nonumber \\
& \Im\left(V_{13}+V_{24}\right) S_1^y - \Im\left(V_{12}+V_{34}\right) S_2^y + \nonumber \\
& 2\Re\left(V_{12}-V_{34}\right) S_1^z S_2^x + 2\Re\left(V_{13}-V_{24}\right) S_1^x S_2^z - \nonumber \\
& 2\Im\left(V_{12}+V_{34}\right) S_1^z S_2^y - 2\Im\left(V_{13}-V_{24}\right) S_1^y S_2^z + \nonumber \\
& 2\Re\left(V_{23}+V_{14}\right) S_1^x S_2^x + 2\Re\left(V_{23}-V_{14}\right) S_1^y S_2^y - \nonumber \\
& 2\Im\left(V_{14}+V_{23}\right) S_1^y S_2^x - 2\Im\left(V_{14}-V_{23}\right) S_1^x S_2^y
\end{align}

We recognize that this can be written in a much more compact form
\be
H_{\rm mag} = \sum_{ab} S_1^a J_{ab} S_2^b + \vec{b}_1 \cdot \vec{S}_1 + \vec{b}_2 \cdot \vec{S}_2
\ee
with
\begin{align}
\vec{b}_1 = \Re\left(V_{13}+V_{24}\right) \hat{x} - \Im\left(V_{13}+V_{24}\right) \hat{y} + \frac{1}{2} \left( V_{11} + V_{22} - V_{33} - V_{44} \right) \hat{z} \\
\vec{b}_2 = \Re\left(V_{12}+V_{34}\right) \hat{x} - \Im\left(V_{12}+V_{34}\right) \hat{y} + \frac{1}{2} \left( V_{11} - V_{22} + V_{33} - V_{44} \right) \hat{z}
\end{align}
and the exchange tensor given by
\be
J = \left( \begin{array}{ccc}
2\Re\left(V_{23}+V_{14}\right) & -2\Im\left(V_{14}-V_{23}\right) & 2\Re\left(V_{13}-V_{24}\right) \\
-2\Im\left(V_{14}+V_{23}\right) & 2\Re\left(V_{23}-V_{14}\right) & -2\Im\left(V_{13}-V_{24}\right) \\
2\Re\left(V_{12}-V_{34}\right) & -2\Im\left(V_{12}+V_{34}\right) & \left( V_{11} - V_{22} - V_{33} + V_{44} \right)
\end{array} \right)
\ee

Finally, we may decompose $J$ into its symmetric $J_S = (J+J^T)/2$ and antisymmetric $J_A = (J-J^T)/2$ parts, the latter of which is entirely responsible for the Dzyaloshinski-Moriya interaction in Eq.~(\ref{eq:exchange}).

Having built the above framework in some generality, one may now insert a particular model to calculate the matrix elements $(H_{\rm mag})_{\alpha\beta}$, and finally generate the exchange matrix and $\vec{b}$ vectors. Thus, inserting
\be
h^{+\hat{x}} = -t
\left( \begin{array}{cc} \cos\alpha & \sin\alpha \\  -\sin\alpha & \cos\alpha \end{array} \right), \quad
h^{+\hat{y}} = -t
\left( \begin{array}{cc} \cos\alpha & i\sin\alpha \\  i\sin\alpha & \cos\alpha \end{array} \right), \quad
\ee
as well as $U_{\su\su}=U_{\sd\sd}=U, \quad U_{\su\sd}=U_{\sd\su}=\lambda U$, we can quickly obtain that the matrix elements conspire such that $\vec{b}_1=\vec{b}_2=0$ along either bond. This should be expected as the underlying model was time-reversal symmetric, so we could have thrown these terms out by hand. The exchange tensors are, respectively,

\be
J^{+\hat{x}} = \frac{4t^2}{gU} \left( \begin{array}{ccc}
 -\cos(2\theta) & 0 & g\sin(2\theta) \\
 0 & -1 & 0 \\
 -g\sin(2\theta) & 0 & (1-2g)\cos(2\theta) \\
\end{array} \right)
\ee

\be
J^{+\hat{y}} = \frac{4t^2}{gU} \left( \begin{array}{ccc}
 -1 & 0 & 0 \\
 0 & -\cos (2 \theta) & -g \sin (2\theta) \\
 0 & g \sin (2 \theta) & (1-2 g) \cos (2 \theta)
\end{array} \right)
\ee
which match precisely the expressions of Eq.~(\ref{eq:exchange}) and the paragraph that follows it.

%%%%%%%%%%%%%%%%%%%%%%%%%%%%%%%%%%%%%%%%%%%%%%%%%%%%%%%%%%%%%%%%%%%%%%%%%%
%% End matter
%%%%%%%%%%%%%%%%%%%%%%%%%%%%%%%%%%%%%%%%%%%%%%%%%%%%%%%%%%%%%%%%%%%%%%%%%%

\bibliographystyle{ws-rv-van}
\bibliography{soclattice}
%\blankpage
%\printindex[aindx]	% to print author index
%\printindex		% to print subject index
\end{document}